\documentclass[aps,pra,twocolumn,floatfix,superscriptaddress,showpacs,amsmath,amssymb]{revtex4-1}
\usepackage{graphicx}
\usepackage{epsfig}
\usepackage{hyperref}
\usepackage{color,colordvi}
\usepackage{float}
\allowdisplaybreaks
\usepackage[disable]{todonotes} % notes not showed
\newcommand{\bea}{\begin{eqnarray}}
\newcommand{\eea}{\end{eqnarray}}
\newcommand{\ba}{\begin{eqnarray*}}
\newcommand{\ea}{\end{eqnarray*}}

\newcommand{\aver}[1]{\langle {#1} \rangle}

\newcommand{\es}[1]{\begin{split}#1\end{split}}
\newcommand{\beq}{\begin{equation}}
\newcommand{\eeq}{\end{equation}}
\newcommand{\beqs}{\begin{equation*}}
\newcommand{\eeqs}{\end{equation*}}
\newcommand{\la}{\left\langle}
\newcommand{\ra}{\right\rangle}

\newcommand{\lp}{\left(}
\newcommand{\rp}{\right)}
\newcommand{\lsq}{\left[}
\newcommand{\rsq}{\right]}
\newcommand{\lbr}{\left\lbrace}
\newcommand{\rbr}{\right\rbrace}
\newcommand{\da}{\dagger}
\newcommand{\bma}{\begin{pmatrix}}
\newcommand{\ema}{\end{pmatrix}}

\newcommand{\bra}[1]{\langle #1 |}
\newcommand{\ket}[1]{| #1 \rangle}
\newcommand{\mo}{{-1}}
\newcommand{\rw}{\rightarrow}
\newcommand{\oh}{\frac{1}{2}}

\newcommand{\w}{\omega}
\newcommand{\pt}{\partial _t}

\newcommand{\ig}[4][]{
\begin{figure}[H]
\centering
\includegraphics[width=#3 \linewidth]{#2}
\caption{#4}\ifthenelse{\equal{#1}{}}{}{\label{#1}}
\end{figure}
}
\newcommand{\re}{\text{Re}}
\newcommand{\im}{\text{Im}}
\newcommand{\tr}{\text{tr}}

\newcommand{\abs}[1]{ \left\lvert #1	\right\rvert}

\usepackage{bbold}
\newcommand{\id}{\mathbb{1}}

\newcommand{\lind}{\hat{\mathcal{L}}}

\begin{document}

\title{Spectral functions and negative density of states of a driven-dissipative nonlinear quantum resonator}

\author{Orazio Scarlatella}
\affiliation{Institut de Physique Th\'{e}orique, Universit\'{e} Paris Saclay, CNRS, CEA, F-91191 Gif-sur-Yvette, France}
\author{Aashish A. Clerk}
\affiliation{Institute for Molecular Engineering, University of Chicago, 5640 S. Ellis Ave., Chicago, IL, 60637, USA}
\author{Marco Schiro}
\affiliation{Institut de Physique Th\'{e}orique, Universit\'{e} Paris Saclay, CNRS, CEA, F-91191 Gif-sur-Yvette, France}

\begin{abstract}
We study the spectral properties of Markovian driven-dissipative quantum systems, focusing on the nonlinear quantum van der Pol oscillator as a paradigmatic example.  We discuss a generalized Lehmann representation, in which single-particle Green's functions are expressed in terms of the eigenstates and eigenvalues of the Liouvillian.  Applying it to the quantum van der Pol oscillator, we find a wealth of phenomena that are not apparent in the steady-state density matrix alone.  Unlike the steady state, the photonic spectral function has a strong dependence on interaction strength.  Further, we find that the interplay of interaction and non-equilibrium effects can result in a surprising ``negative density of states", associated with a negative temperature, even in absence of steady state population inversion.

\end{abstract} 

\maketitle
%%%%%%%%%%%%%%%%%%%%%%%%%%%%%%%%%%%%%%%%%%%%%%%%%%%%%%%%%%%%
%%%%%%%%%%%%%%%%%%%%%%%%%%%%%%%%%%%%%%%%%%%%%%%%%%%%%%%%%%%%

\clearpage

% \comm{introduction and conclusion are still to be done, you can skip them (or write something if you want). I left some comments to highlight where I put the things we discussed about last time and to highlights some doubts I have. Have a good reading and thank you!}

\section{Introduction}
Recent experimental progress in controllable quantum systems has renewed the interest in driven-dissipative quantum phenomena.  Such systems typically have non-trivial, non-thermal steady states determined by the balancing of drive and dissipation.  Examples include atomic and optical systems such as ultracold gases in optical lattices~\cite{BlochDalibardNascimbeneNatPhys12} or trapped ions~\cite{BlattRoosNatPhys12}, as well as solid state systems such as arrays of nonlinear superconducting microwave cavities~\cite{Wallraf_Nature2004,AndrewNatPhys,LeHurReview16} or microcavity polaritons realizing quantum fluids of light~\cite{carusotto_quantum_2013}. The simplest regime to consider is where dissipative effects are describable by a standard Markovian Lindblad master equation.  Even here, considerable complexity can arise if there are interactions (nonlinearities).  A vast amount of theoretical work has focused on finding (either exactly or approximately) the steady state of such systems, and the corresponding steady-state expectation values of observables~\cite{OrusVidalPRB08,FinazziEtAlPRL15,JinEtAlPRX16,LiPetruccioneKochPRX16,
NohAngelakisRepProgPhys2016}.

While describing steady states is clearly of interest, many experimental probes involve studying how a system responds to a weak applied perturbation.  One is then naturally interested in understanding the Green's functions that describe the linear response of the system to external perturbations.  For Markovian systems, these correlations functions can be readily computed using the quantum regression theorem, and have been studied in a variety of different contexts, from the standard example of resonance fluorescence of a driven two-level atom~\cite{MollorPhysRev69,Astafiev840,CarrenoEtAlLasPhot17}, recently discussed in the case of arrays of coupled qubits~\cite{KildaKeelingArxiv17}, to the second-order correlations probing bunching/anti-bunching of time-delayed photons (see, e.g.,~\cite{LangEtAlPRL11}). The topic of correlation functions is also a standard topic in almost any quantum optics textbook (see, e.g.,~\cite{carmichaelStatistical1999,breuerPetruccione2010}).

Despite this existing work, methods for obtaining physical intuition from the behaviour of correlation functions (or their corresponding spectral functions) remain of interest.  For closed, equilibrium quantum many body systems, the Lehmann representation~\cite{Kallen:1952zz,Lehmann1954} (see also, e.g.,\cite{altlandSimons2010}) is a powerful tool.  It expresses a single particle spectral function in terms of the system energy eigenstates, and allows one to interpret the spectral function in terms of Fermi Golden rule rates for the addition (or removal) of a particle.  This directly connects to experimental probes (e.g.~ARPES or tunneling spectroscopy), and is invaluable in constructing intuitive pictures.
\begin{figure}[t]
\begin{center}
\epsfig{figure=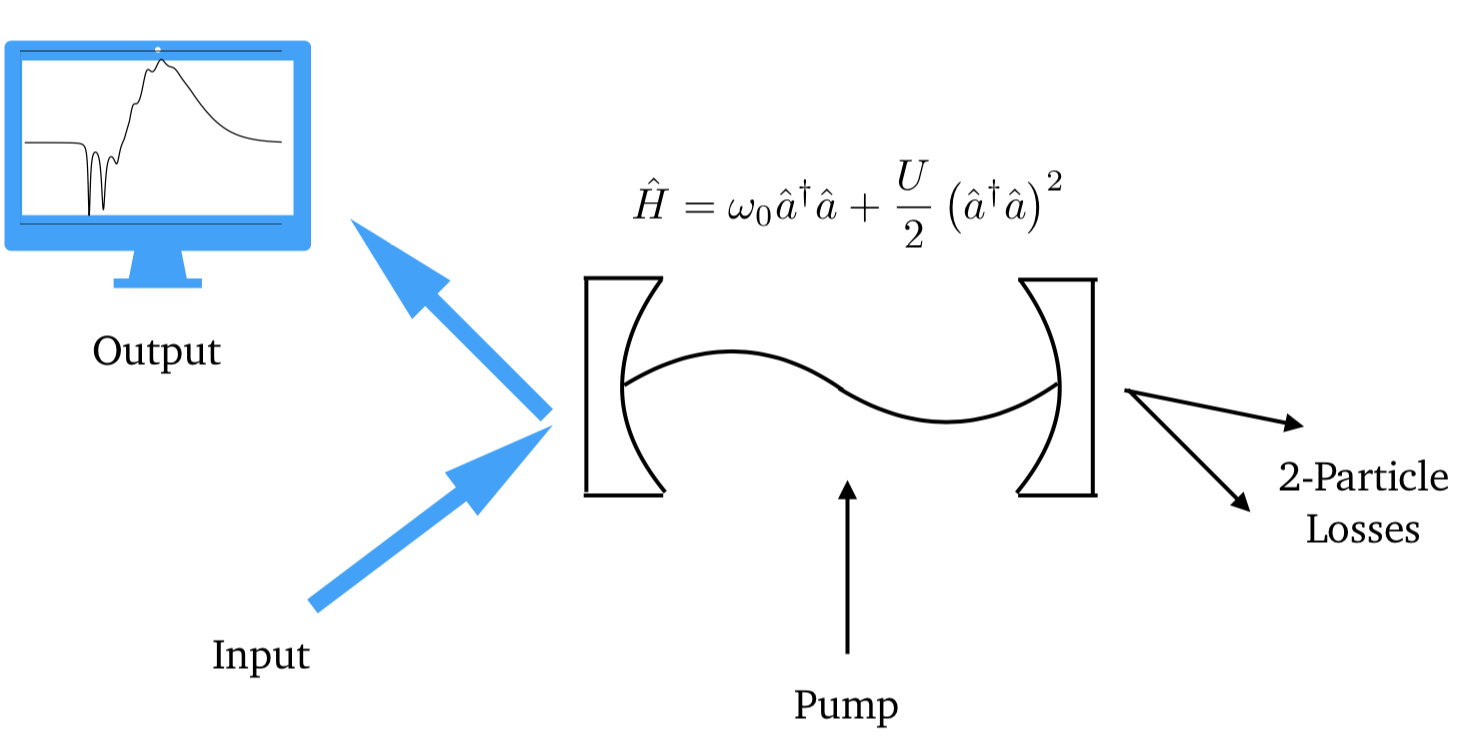,scale=0.3}
\caption{Schematic plot of the setup considered in this manuscript. A cavity mode with Kerr nonlinearity is driven by an incoherent pump and subject to two-photons losses. We investigate its spectral features, encoded in the cavity mode spectral function.  This quantity could be directly measured by considering the reflection of a weak probe tone.}
\label{fig:image}
\end{center}
\end{figure}

In this work, we formulate the corresponding Lehmann representation of correlations functions of a driven dissipative system, and show that it also serves as a powerful interpretive tool.  We focus on systems described by a Markovian Lindblad master equation, and discuss how the spectral functions are directly connected to the dynamical modes of the corresponding Liouvillian.  As a concrete example, we analyze a simple but non-trivial model of driven-dissipative nonlinear quantum van der Pol oscillator, describing a single-mode bosonic cavity with a Kerr interaction subject to incoherent driving and nonlinear loss (see Ref.~\onlinecite{DykmanKrivoglaz84} for a comprehensive review). This model has recently received attention in the context of quantum synchronization~\cite{LeeSadeghpourPRL13,WalterNunnekampBruderPRL14,LorchEtAlPRL16}; it is also directly realizable in superconducting circuit architectures, where strong Kerr interactions and engineered two-photon losses have been experimentally achieved~\cite{Leghtas2015,TouzardEtAlPRX18}.
% This model is particularly useful since its steady state density matrix has a remarkably simple structure, which is insensitve to coherent interactions and only set by drive and dissipation. 
While the model has a relatively simple steady state, its spectral features are instead remarkably rich~\cite{dykman1978,DykmanKrivoglaz84}. Unlike the steady state, the spectral function depends strongly on the size of the Kerr interaction, and reveals physics beyond that in the steady state density matrix.  Specifically we show that the model features both population inversion in the density matrix and a negative density of states (NDoS), two aspects which are tightly connected in equilibrium but whose interplay in the driven-dissipative case appears to be more complex. In particular we find a regime where NDoS emerges, even in absence of a population inversion in the stationary density matrix.
%We discuss how the structure of the spectral function here can be directly tied to standard $T_1$ and $T_2$ relaxation processes.  Further, we show that the inversion of the population in the steady state only manifests itself in the spectral functions for sufficiently large interaction strength.

The paper is organized as follows. In Sec.~\ref{sectGF} we define
Greens' functions for an open system governed by a Lindblad master equation and discuss their decomposition in terms of eigenstates of the Liouvillian. In Sec.~\ref{sectModel} we introduce the specific model of a driven-dissipative Kerr resonator, and in Sec.~\ref{sectResults} we discuss its spectral properties. We conclude in Sec.~\ref{sectConcl}.
\section{Basic results}
\label{sectGF}

\subsection{Lindblad dynamics and Liouvillian spectrum}

We consider an open quantum system described by the Lindblad master equation 
%(say smth on fermions)
\beq
\label{eq:lindblad}
\pt \hat{\rho}(t)  = \lind ( \hat{\rho}(t) )
\eeq
where $\hat{\rho}$ is the reduced density matrix of the system and $\lind$ is the Lindblad/Liouvillian super-operator which takes the general form ($\hbar = 1$)
\beq
\lind ( \hat{\rho })=-i[\hat{H},\hat{\rho}]+\sum_{\alpha}\hat{L}_{\alpha}\hat{\rho }\hat{L}^{\dagger}_{\alpha}-\frac{1}{2}
\lbr \hat{L}^{\dagger}_{\alpha} \hat{L}_{\alpha} , \hat{\rho }\rbr
\eeq
The first term on the RHS describes (unitary) Hamiltonian evolution, whereas the remaining terms describe incoherent driving and dissipative processes (each corresponding to a time-independent operator $\hat{L}_{\alpha}$).  Note that we will use throughout calligraphic letters to indicate super-operators.

%%%%%%
We assume the most common case where 
Eq.~(\ref{eq:lindblad}) has a unique time-independent stationary state $\hat{\rho}_s$.
%\add{ref spohn}.  
We are interested in calculating two-time correlation functions, which depend only on time differences due to the time-translational invariance of the stationary state. Under the same assumptions one makes to derive Eq. \eqref{eq:lindblad}, one can write dynamical correlators in terms of the Liouvillian, a result known as the quantum regression formulae \cite{carmichaelStatistical1999,breuerPetruccione2010}. For example, assuming $t>0$, we have that 
%We make the example of the retarded parts (non zero for $t>0$ only) of the greater and lesser single particle Green's functions \add{they are the minimal building block for all the Green's functions...because with semigroups we can compute only retarded functions..} defined as follows: 
\begin{align}
\label{eq:greatGreenFunc}
 \la \hat{A}(t) \hat{B } (0) \ra  =  \tr \lp \hat{A} \, e^{t \lind } \lp {\hat{B}} \hat{\rho}_s \rp \rp  
\end{align}
where $\hat{A}(s),\hat{B}(s)$ are generic Heisenberg-picture operators of the system.
%$a$, $a^\da$ are the bosonic annihilation/creation operators 
%and we have assumed $\tr \rho_s =1 $.
%\add{say what does this Green's function mean: one particle excitation}
%%%

Recall that for closed systems in thermal equilibrium, it is extremely useful to relate Green's functions directly to the energy eigenvalues and eigenstates of the system; this is achieved by the Lehmann representation \cite{bruusFlensberg2004,stefanucciVanLeeuwen2013}. Our goal is to do something analogous in our driven dissipative system.  Now however, the relevant spectrum is not that of the system Hamiltonian $\hat{H}$, but rather that of the Liouvillian.  Such spectral decompositions for correlations functions for Lindblad open systems have been derived before (see e.g.~\cite{arrigoniWolfgangPRL2013,DordaEtAlPRB14} in the context of electron transport through correlated impurities or Ref.~\onlinecite{DykmanKrivoglaz84} in the contex of non-linear oscillators where a  related decomposition as sum of partial spectra was introduced), but we include it here again for completeness and clarity.

To proceed, the first step is to properly enumerate the the eigenmodes of the Liouvillian.  
It is useful to treat the space of operators (so-called Liouville space) as a Hilbert space, with the inner product $\aver{ \hat{A},\hat{B}} \equiv \tr \lp \hat{A}^\da \hat{B }\rp $ \cite{albertThesisArxiv2018}. This allows one to define the adjoint Liouvillian by $\aver{ \hat{A},\lind ( \hat{B} ) } = \aver{\lind^\da (  \hat{A} ),  \hat{B}}$.  The 
left eigenvectors ($\hat{l}_\alpha$), right eigenvectors ($\hat{r}_\alpha$), and eigenvalues $\lambda_\alpha$
of the Liouvillian are then defined via 
\begin{eqnarray}
	\lind \lp \hat{r}_\alpha \rp = \lambda_\alpha \hat{r}_\alpha \\ 
	\lind^\da \lp \hat{l}_\alpha \rp = \lambda_\alpha^* \hat{l}_\alpha
\end{eqnarray}  
Note that $\hat{l}_\alpha$ and $\hat{r}_\beta$ are mutually orthogonal: $\tr\lp \hat{l}_\alpha^\dagger \hat{r}_\beta \rp =N \delta_{\alpha,\beta} $.  For convenience, we fix the normalization constant $N=1$, such that we have a simple completeness relation 
\beq \label{eq:completeness} 
	\sum_\alpha \hat{r}_\alpha \hat{l}_\alpha^\da = \hat{\id} 
\eeq

The eigenmodes of the Liouvillian directly determine how the system relaxes to the steady state.  First, note that the unique (by assumption) steady state of our system corresponds to the unique right eigenstate of $\lind$ with zero eigenvalue. We label this eigenstate by the index $\alpha = 0$, thus 
$\hat{\rho}_s = \hat{r}_0 / \tr \lsq \hat{r}_0 \rsq$ and $\lambda_0 = 0$. 
Suppose now that at $t=0$ the system starts in some state $\hat{\rho}(0)$ that is not the stationary state.  At later times, the system reduced density matrix will be given by
\beq
	\label{eq:decayModes} 
	\hat{\rho}(t) - \hat{\rho}_s =  \sum_{\alpha \neq 0} 
    c_\alpha 
%     e^{(\re \lambda_\alpha + i \im \lambda_\alpha)t} 
    e^{\lambda_\alpha t} 
    \hat{r}_\alpha 
\eeq
with
\begin{equation}
    c_{\alpha} = \tr \left( \hat{l}_\alpha^\dagger \hat{\rho}(0) \right).
\end{equation}
We can thus interpret each Liouvillian eigenmode $\alpha$ as a possible dynamical decay mode of some initial deviation from the steady state, with a decay rate given by $-\re \, \lambda_\alpha$.  In general, a given decay mode will involve both diagonal elements of the density matrix in the energy-eigenstates basis (i.e.~populations) as well as off-diagonal elements (i.e.~coherences).  However, in some cases the situation simplifies and one can cleanly separate the eigenmodes into processes only involving populations ($T_1$ processes) or only involving coherences ($T_2$ processes); we will see this explicitly in Sec.~\ref{subsect:sym}.   

%%%%%%%%%%%%%%%%%%%%%%%%%%
\subsection{Spectral representation of correlation functions}

To derive the Lehmann representation of the correlation function in 
Eq.~\eqref{eq:greatGreenFunc}, we note that the operator $\hat{B}$ acting on steady state density matrix $ \hat{\rho}_s $ causes the system to deviate from the steady state.  Just as in Eq.~(\ref{eq:decayModes}), this deviation can be expressed as a linear combination of the Liouvillian decay modes (i.e.~in terms of the right- eigenstates of the Liouvillian).  We thus obtain:  
\beq
	\label{eq:lehmannGreat}
 	\la \hat{A}(t) \hat{B} (0) \ra = 
    \sum_\alpha e^{ \lambda_\alpha  t } \, \tr \lp  \hat{A} \hat{r}_\alpha \rp  \tr \lp \hat{l}_\alpha^\da \hat{B}\hat{\rho}_s \rp.
\eeq
% where the coefficients of the expansion are given by the projection on left-eigenstates. 
At an intuitive level, $\hat{B}$ ``excites" the various dynamical eigenmodes of the Liouvillian; these modes then oscillate and decay as a function of time.  The factor involving $\hat{A}$ corresponds to the change in $\left \langle \hat{A} \right \rangle$ (compared to the steady state value) associated with exciting a particular dynamical eigenmode $\alpha$.
% Each term of the sum corresponds to a decay mode of the excitations performed by operators $\hat{A}$ and $\hat{B}$, with lifetime $\re{\lambda_\alpha}$ as we discussed for Eq. \eqref{eq:decayModes}.
% \new{The terms$ \tr [  \hat{A} \hat{r}_\alpha ] $($\tr [ \hat{l}_\alpha^\da \hat{B} \hat{\rho}_s ]$)  measures how much the operator $\hat{A}^\da$($\hat{B} \hat{\rho_s}$) overlap with the right(left) eigenstate $\hat{r}_\alpha$($\hat{l}_\alpha$). If, for example,   $\hat{A}^\da$  is orthogonal to $\hat{r}_\alpha$, then the corresponding term in the sum vanishes. 
In what follows, we will focus on the connected average $ \la \hat{A}(t) \hat{B} (0) \ra -  \la \hat{A}(t) \ra \la \hat{B} (0) \ra$; equivalently, we will shift $\hat{A}$ and $\hat{B}$ each by a constant so that they vanish in the steady state.  In this case, the steady state (i.e.~$\alpha = 0$) does not contribute to the sum in Eq.~(\ref{eq:lehmannGreat}).

%\AC{Issue: modified discussion here to make it clear that taking the zero dissipation limit is subtle, as in this case there is no a priori unique steady state.}

It is interesting to see how one recovers the standard closed-system Lehmann represetation by taking the zero-dissipation limit of Eq.~(\ref{eq:greatGreenFunc}).
This limit implies only keeping the Hamiltonian term in Eq.~(\ref{eq:lindblad}), i.e.~replacing $\lind$ with $-i\lsq \hat{H}, \bullet \rsq$ (where $\bullet$ is the operator on which $\lind$ acts).  Letting $\ket{\psi_i}$ and $E_i$ denote the eigenstates and eigenvalues of the Hamiltonian $\hat{H}$, it is straightforward to find the dynamical eigenmodes of $\lind$.  Each dynamical eigenmode $\alpha$ corresponds to a pair of energy eigenstates $i,j$:
\begin{eqnarray}
	\label{eq:eigstatZeroOrd}
	\hat{r}^{(0)}_{i,j} = \hat{l}	^{(0)}_{i,j} = \ket{\psi_i} \bra{\psi_j} \\ 
	\label{eq:eigvalZeroOrd}
	\lambda_{i,j}^{(0)}= - i \lp E_i - E_j \rp.
\end{eqnarray}
These modes have a simple interpretation.  For a closed system, populations in the energy eigenstate basis are time-independent, corresponding to the zero-eigenvalue modes $\lambda_{i,i}^{(0)}$.  Further, the coherences in the energy eigenstate basis have a simple undamped oscillatory behaviour, corresponding to the $i \neq j$ modes. 
We stress that in the purely closed system case, the dynamics no longer picks out a unique steady state, as any incoherent mixture of energy eigenstates
is stationary. The only constraint from the dynamics to gaurantee stationarity is that $\hat{\rho}_s$ be diagonal in the energy eigenstate basis:  $\hat{\rho}_s = \sum_k p_k \ket{\psi_k} \bra{\psi_k}$. As usual, one must then assume the probabilities $p_k$ when computing average values and correlation functions.   % This is required for stationarity.  
Formally, this assumption corresponds to the usual limit where the dissipation is non-zero but infinitesimally weak; in this limit, dissipation can determine the steady state, but does not impact dynamics.

Using the above eigenmodes in Eq.~\eqref{eq:lehmannGreat} and defining $E_{ij} = E_i - E_j$, we obtain
\begin{eqnarray}
	\label{eq:lehmannGreatClosed}
% 	\es{
 	\la \hat{A}(t) \hat{B }(0) \ra  
    &= \sum_{ij} e^{- i E_{ij} t } \bra{\psi_j}  \hat{A}  \ket{\psi_i}  \bra{\psi_i} \hat{B}  \hat{\rho}_s \ket{\psi_j} 
% }
\nonumber \label{eq:greatLehEq} \\
%\mathcal{F} \lp G^>(t)\theta(t)  \rp (\w)= \sum_{ij}\frac{  \abs{ \bra{\psi_i} a ^\da \ket{\psi_j} }^2 p_j}{\w -  \lp E_i - E_j \rp + i \eta }
%  \la \hat{A}(t) \hat{B} (0) \ra 
 &= \sum_{ij}\, e^{- i E_{ij} t } \bra{\psi_j}  \hat{A  }\ket{\psi_i} \bra{\psi_i} \hat{B}\ket{\psi_j}  p_j 
\end{eqnarray}
The first line matches what one would obtain from a direct calculation using
$ \la \hat{ A}(t) \hat{B} (0) \ra =  \tr \lp e^{i \hat{H}t}\hat{A} e^{-i\hat{H}t} {\hat{B}} \hat{\rho}_s  \rp$.  In the second line, we have used the diagonal form of $\hat{\rho}_s$.  For a system in thermal equilbrium, the $p_k$ are simple Boltzmann weights; we then recover the usual textbook thermal equilibrium formula (see, e.g.,
\cite{bruusFlensberg2004,altlandSimons2010}).

\subsection{Single particle Green's functions}	
\label{subsectGF}

In the rest of the paper we will focus on the retarded single-particle Green's function of a bosonic system.  Letting 
$\hat{a}$ denote the canonical bosonic annihilation operator, the single-particle Green's function is defined as
\beq 
	\label{eq:defRetared}
	G^R(t) = - i \theta(t)\aver{ \lsq \hat{a}(t), \hat{a}^\da (0) \rsq } 
\eeq 

This correlation function plays an important role in many different contexts.
For example, via the Kubo formula, it describes the linear response of the expectation $\aver{\hat{a}(t)}$ to a weak, classical field $h(t')$, which couples linearly to $\hat{a}^\da$.  
In the case where $\hat{a}$ describes a photonic cavity mode, $G^R(t)$ can be directly measured by weakly coupling the cavity to an input-output waveguide and measuring the reflection of a weak probe tone (see e.g.~\cite{LemondePRL2013,lavitanClerkNJP2016}).

% In the rest of the paper we will focus on retarded Green's functions. These determine the linear change of the average value of a an operator $\aver{\hat{A}}$ when the system is perturbed by an external field coupling to $\hat{B}$ through the Kubo formula \cite{stefanucciVanLeeuwen2013}.  We will pay particular attention to the single-particle Green's function of a bosonic system, implying that we take $\hat{A}=\hat{a}$, $\hat{B}=\hat{a}^{\dagger}$ where $\hat{a}$ is the canonical annihilation operator for a particular state (satisfying $[\hat{a},\hat{a}^{\dagger}]=1$). The retarded single particle Green's function is defined as

\subsubsection{Closed stationary system}
\label{subsect:GFclosed}
For a closed system in a time-independent steady state, the Fourier transform of the Lehmann representation of the retarded Green's function is \cite{altlandSimons2010}
\begin{align}
	G^R(\w) 
%     & =  
%     \sum_{i,j} p_i
%     \left(
%     \frac{  \abs{\bra{\psi_j } \hat{a}^\da \ket{\psi_i}}^2   }
%     {  \w - E_j + E_i  + i \eta} 
% -    \frac{  \abs{\bra{\psi_j } \hat{a} \ket{\psi_i}}^2   }
%     {  \w - E_i + E_j  + i \eta} 
% \right) \nonumber \\
	& =  \sum_{i,j} 
    \frac{  \abs{\bra{\psi_j } \hat{a}^\da \ket{\psi_i}}^2 \lp p_i   - p_j \rp  }
    {  \w - E_j + E_i  + i \eta}
    	\label{eq:lehRetEq1}
\end{align}
where $\eta$ is a positive infinitesimal.  Consistent with causality, this function is analytic in the upper half plane.  It has simple poles with infinitesimal negative imaginary part, and with purely real weights. 
% The corresponding spectral function $A[\omega]$ has the form:
% \begin{align}
% 	A(\w) & =  \sum_{i,j} p_i
%     \Big(
%     \abs{\bra{\psi_j } \hat{a}^\da \ket{\psi_i}}^2   
%     \delta \left( \w - E_j + E_i \right) \nonumber \\
%     &
% 	-      \abs{\bra{\psi_j } \hat{a} \ket{\psi_i}}^2   
%       \delta(\w - E_i + E_j) \Big)
% \end{align}
% $A(\omega)$ has a simple interpretation in terms of Golden rule transition rates.  The first term is naturally associated with adding a particle to the steady state and creating an excitation with energy $\omega$, whereas the second term is associated with removing a particle and creating an excitation with energy $-\omega$.   
Of particular interest is the imaginary part of $G^R(\omega)$, which defines the  single-particle spectral function or density of states $A(\omega)$:
\beq 
	\label{eq:specFunc} 
    A(\w)= - \frac{1}{\pi} \, \im G^R(\w). 
\eeq
%This plays the role of an effective single-particle density of states, and can be directly probed using a variety of different experimental tools.  For example, in the case where $\hat{a}$ describes a photonic cavity mode, the spectral function can be measured by weakly coupled the cavity to an input-output waveguide and measuring the reflection of a weak probe tone (see e.g.~\cite{lavitanClerkNJP2016}).

% $ A(\w)$ satisfies the sum rule
% \beq
%  \int_{-\infty}^{\infty} d\w A(\w) = \aver{\lsq \hat{a}, \hat{a}^\da \rsq} = 1
% \eeq
% One can verify it from Eq. \eqref{eq:lehmannRet}, recalling that Lorentzians have unit area, $\int_{-\infty}^\infty d \w \frac{1}{\pi}\frac{\Gamma}{(\w + \Lambda)^2 + \Gamma^2} =1 $, and upon using the completeness relation in Eq. \eqref{eq:completeness}. As a result, interactions and dissipation can reshape $A(\w)$, but they cannot change its area.
For a closed system, the spectral function follows directly from Eq.~(\ref{eq:lehRetEq1}): 
 \begin{align}	
 	A(\w) & =  \sum_{i,j} p_i
     \Big(
     \abs{\bra{\psi_j } \hat{a}^\da \ket{\psi_i}}^2   
     \delta \left( \w - E_j + E_i \right) \nonumber \\
     &
 	-      \abs{\bra{\psi_j } \hat{a} \ket{\psi_i}}^2   
       \delta(\w - E_i + E_j) \Big) \nonumber \\
     & = 
     \sum_{i,j} (p_i-p_j)     
     \abs{\bra{\psi_j } \hat{a}^\da \ket{\psi_i}}^2   
     \delta \left( \w - E_j + E_i \right)
     \label{eq:ClosedSpectralFunc}
 \end{align}
The first equality allows us to give a simple physical interpretation of $A(\omega)$ in terms of Golden rule transition rates.  The first term is naturally associated with adding a particle to the steady state and creating an excitation with energy $\omega$, whereas the second term is associated with removing a particle and creating an excitation with energy $-\omega$.   

The second equality in Eq.~(\ref{eq:ClosedSpectralFunc}) also leads to an important result.  If we assume that $p_j \leq p_i$ whenever $E_j \geq E_i$, then we immediately can conclude:
\begin{align}
	\label{eq:signPropA} 
	&A(\w) \gtrless 0  &\text{for } \w \gtrless  0, 
\end{align}
i.e.~the sign of the spectral function $A(\omega)$ matches the sign of $\omega$.  A violation of this condition indicates the existence of population inversion in the steady state:  a higher-energy energy eigenstate has a larger population in the steady state than a lower-energy eigenstate.  While this is impossible in thermal equilibrium, it is indeed possible in a generic driven-dissipative non-thermal steady state.  We will discuss the emergence of population inversion in the spectral function in great detail in the next section, in the context of a specific system.  

\subsubsection{Open system case}

We can use the generalized Lehmann representation, Eq.~(\ref{eq:lehmannGreat}), to derive a corresponding result for the single-particle retarded Green's function of a Linbdlad open system.  We obtain
\beq
	\label{eq:lehmannRet}
	G^R(\w) =  
    	\sum_\alpha 
        \frac{ w_{\alpha} } 
	{\w + \im \lambda_\alpha - i \re \lambda_\alpha } 
\eeq
% \beq
% 	\label{eq:lehmannRet}
% 	G^R(\w) =  \sum_\alpha \frac{\tr \lsq \hat{a} \hat{r}_\alpha \rsq\lp 
%     \tr \lsq \hat{l}_\alpha^\da \hat{a}^\da \hat{\rho}_s  \rsq  
%     - \tr \lsq \hat{l}_\alpha^\da \hat{\rho}_s \hat{a}^\da \rsq \rp}
% 	{\w + \im \lambda_\alpha - i \re \lambda_\alpha } 
% \eeq
with $w_{\alpha}=\tr \lp \hat{a} \hat{r}_\alpha \rp  \tr \lp \hat{l}_\alpha^\da [ \hat{a}^\da, \hat{\rho}_s ]  \rp$.
There is clearly some similarity to the closed-system expression Eq.(\ref{eq:lehRetEq1}).  Like the closed-system case, the Green's function is decomposed into a sum of simple poles.  However, whereas for the closed system poles occurred at energy differences that were infinitesimally shifted from the real axis, now the poles occur at eigenvalues of the Liouvillian, and will be shifted a finite distance below the real axis.  

More intriguingly, the residues $w_\alpha$ associated with the poles of $G^R(\omega)$  are no longer necessarily real (as it must be for a closed system).  %This ultimately arises from the non-hermitian nature of the Lindbladian generating the dynamics: since right and left eigenmodes are not simply related by complex conjugation, the amplitude for creating or removing a particle from these eigenmodes does not differ only by a phase. 
This has a direct consequence on the spectral function (c.f. Eq.~(\ref{eq:specFunc})), which now takes the form
%A direct consequences of this observation arises when considering the spectral function, defined as previously as $ A(\w)=-1/\pi\mbox{Im}G^R(\omega)$, which reads
\beq\label{eqn:LehmannA}
A(\w)=-\frac{1}{\pi}\sum_{\alpha}
    z_{\alpha}\left(\w\right) \frac{ \left(\re\lambda_\alpha\right)}
{\left(\omega+\im\lambda_{\alpha}\right)^2+\left(\re\lambda_\alpha\right)^2}
\eeq
where 
\beq
    z_{\alpha}\left(\w\right)=\re w_\alpha+
    \im w_{\alpha}
    \frac{\omega+\im\lambda_\alpha }{\re\lambda_\alpha}.
\eeq
It follows that the spectral function is no longer simply a sum of Lorentzians.  An immediate corollary is that unliked the closed-system case (c.f.~Eq.(\ref{eq:signPropA})),  the sign of the spectral function is not controlled in a simple way by the structure of the steady state distribution. In other words, a driven-dissipative systems can have spectral functions which violate the sign property Eq.~\eqref{eq:signPropA}, without this necessarily coming from an inverted population of the stationary state.  We will see this explicitly in Sec.~\ref{sectResults} in the context of a driven quantum Kerr cavity. 

We remark that the spectral function in Eq.~\eqref{eqn:LehmannA} satisfies the sum rules originating from the commutation relations of operators at equal time:
%. In particular we have  
 \beq
 \label{eq:sumProp}
  \int_{-\infty}^{\infty} d\w A(\w) = \aver{\lsq \hat{a}, \hat{a}^\da \rsq} = 1,
 \eeq
 as one can directly verify from Eq. \eqref{eq:lehmannRet}.
 %, recalling that Lorentzians have unit area, $\int_{-\infty}^\infty d \w \frac{1}{\pi}\frac{\Gamma}{(\w + \Lambda)^2 + \Gamma^2} =1 $, and upon using the completeness relation in Eq. \eqref{eq:completeness}. 
 As a result, interactions, driving and dissipation can reshape $A(\w)$, but they cannot change its area. 
 
Our spectral decomposition makes it clear that one can extract information on the eigenvalues of the Liouvillian from the frequency dependence of $G^R(\omega)$.  This could be particularly useful in extended systems, e.g., to determine a dissipative phase transition in which the eigenvalue $\lambda_{\alpha}$ with smallest non-zero real part becomes purely imaginary \cite{kesslerLukinCiracPRA2012,MingantiEtAlPRA18,scarlatellaEtAlArxiv2018}.  
We note that in general it is difficult to compute the spectrum of a Liouvillian, whereas the Green's function may be found via many body techniques~(see, e.g.,\cite{SiebererRepProgPhys2016}).
%Such a divergent susceptibility is a hallmark of a phase transition (though typically the divergence happens at strictly zero frequency).  
%Finally, we notice that the spectral function plays the role of an effective single-particle density of states, and can be directly probed using a variety of different experimental tools \cite{LemondePRL2013,lavitanClerkNJP2016}.

\subsection{Effective temperature}
\label{subsec:effTemp}
%
%\AC{As we don't define a general retarded GF, I made all the discussion here specific to the single particle GFs.}

%We conclude this section by introducing a notion of out-of-equilibrium temperature which has a simple operational meaning and which is directly related to the spectral function $A(\w)$.
As we have seen, Green's functions (via the spectral function $A[\omega]$) can provide information on the effective single-particle density of states of our system.  They can also provide information on how these states are occupied, i.e.~the effective distribution function or the effective temperature of our system.  To obtain this information, one must consider the Keldysh single-particle Green's function:
\beq
    G^{K}(t-t') = -i \aver{ \lbr \hat{a}(t),\hat{a}^\dagger(t') \rbr }
\eeq
which heuristically  describes the fluctuations of the observable $\hat{a}$.  If the system was in true thermal equilibrium, the quantum fluctuation-dissipation theorem (FDT) would have required 
\cite{kamenev2011field,stefanucciVanLeeuwen2013}
\beq
\label{eq:fdt}
\frac{G^{K}(\w)}{-2 \pi i A(\w)} = \coth \lp \frac{ \w }{2 T } \rp 
\eeq
%where $G^{K}(\w)$ is the Fourier transform of the Keldysh correlation function, which for bosonic, time-translational invariant systems is defined as
%\beq
%G^{K}(t-t') = -i \aver{ \lbr \hat{A}(t),\hat{B}(t') \rbr }
%\eeq
where $T$ is the system temperature.

In a non-equilibrium system, there is no well-defined temperature and the FDT does not hold in general.  Nonetheless, it is useful to use the LHS of the FDT relation in Eq.~(\ref{eq:fdt}) to {\it define} at each frequency an effective temperature $T_{\rm eff}(\omega)$, i.e.:
\beq
    \label{eq:effTemp}
    \frac{G^{K}(\w)}{-2 \pi i A(\w)} \equiv \coth \lp \frac{ \w }{2 T_{\text{eff}}(\w) } \rp 
\eeq
As discussed extensively in Ref.~\cite{ClerkRMP2010}, this $T_{\rm eff}[\omega]$ has a direct operational meaning and is a useful quantity in many different physical contexts (e.g.~the theory of optomechanical cavity cooling using driven resonators \cite{MarquardtPRL2007}).  In general, if a second narrow-bandwidth auxiliary system interacts weakly with our main system via exchanging photons, it will equilibrate to a temperature $T_{\rm eff}(\omega_{\rm aux})$, where $\omega_{\rm aux}$ is the frequency of the auxiliary system.  As example, the auxiliary system could be a qubit with splitting frequency $\omega_{\rm aux}$, which interacts with the main system via $H_{\text{int}} \propto \lp \hat{\sigma}_+ \hat{a} + \text{h.c.} \rp$ \cite{ClerkRMP2010,lavitanClerkNJP2016}.  
%Systems in out-of-equilibrium stationary states do not satisfy FDT and the ratio $G^{K}(\w)/ \lp - 2 \pi i A(\w) \rp$ also depends on the specific choice of the operators defining the Green's functions.
%Nevertheless, it is natural to characterize these states by drawing connections to equilibrium, in particular defining effective temperatures. 
%There is no unique way to define an effective temperature and its physical meaning strongly depends on its definition. 
%Here we generalize Eq.~\eqref{eq:fdt} for single particle Green's functions, $\hat{A}=\hat{a}$, $\hat{B}= \hat{a}^\da$, by introducing the frequency-dependent effective temperature $T_{\text{eff}}(\w)$: 
%$T_{\text{eff}}(\w)$ has a simple operational meaning  \cite{lavitanClerkNJP2016}: if a cavity described by $T_{\text{eff}}(\w)$ were to be weakly coupled by an Hamiltonian $H_{\text{int}} \propto \lp \hat{\sigma}_+ \hat{a} + \text{h.c.} \rp$ to a qubit with splitting frequency $\w_c$, this would reach the equilibrium steady state at temperature $T_{\text{eff}}(\w_c)$.
%
%\textcolor{green}{why is the keld purely im ?}
%\AC{Discussed in Kamanev; can also prove it by defining frequency-resolved operators $\hat{a}[\omega]$, $\hat{a}^\dagger[\omega]$.}

Note that for a general non-equilibrium system, there is no requirement that the effective temperature $T_{\rm eff}[\omega]$ be positive.  The fact that $i G^K(\w) > 0$ implies that the sign of $A(\w)$ dictates that of  $T_{\text{eff}}(\w)$. In particular, if the sign of $A(\w)$ obeys the equilibrium property Eq.~\eqref{eq:signPropA}, then $T_{\text{eff}}(\w) > 0 $ $\forall \w$.  If this is not true, then there will be frequency regions in which $T_{\text{eff}}(\w)$ is negative.  We thus see that the anomalous sign of the spectral function discussed earlier is directly connected to the existence of negative effective temperatures.  We stress that this negative temperature has physical consequences.  Again, consider weakly coupling an auxiliary qubit to our system.  If the qubit splitting frequency $\w_\text{aux}$ is such that $T_{\text{eff}}(\w_\text{aux}) <0$, the qubit would thermalize at negative temperature, implying a population inversion (i.e.~higher probability for the qubit to be in the excited state rather than its ground state).

Finally, we stress that in general, $T_{\rm eff}[\omega]$ for a non-equilibrium system will both be frequency dependent {\it and} operator dependent.  That is, if one defined $T_{\rm eff}[\omega]$ using the FDT relation for Green's functions corresponding to operators other than $\hat{a}, \hat{a}^\dagger$, one would in general obtain a different function $T_{\rm eff}[\omega]$ \cite{ClerkRMP2010,KildaKeelingArxiv17}.

\section{Application to the Driven-Dissipative Van der Pol Oscillator}\label{sectModel}

We now use the general results of the previous section to study a specific, non-trivial driven dissipative system.  We consider the nonlinear version of the well-known quantum van-der Pol oscillator \cite{DykmanKrivoglaz84,LorchEtAlPRL16}:  
a bosonic mode with a Kerr nonlinearity subject to incoherent single-particle driving, and two particle losses.  It is described by the master equation ($\hbar=1$)
\begin{align}
\label{eq:masterEquation}
\partial_t \hat{\rho} &= - i [\hat{H},\hat{\rho}] + \gamma \lp r \mathcal{\hat{D}}_p^{(1)} + \mathcal{\hat{D}}^{(2)}_l  \rp [\hat{\rho}]\\
\hat{H} &= \w_0 \hat{n}  +\frac{U}{2} \hat{n}^2 \label{eq:hamiltonian} \\
\mathcal{\hat{D}}_p^{(1)}[\rho] &=  \, \hat{a}^\da \hat{\rho} \hat{a} - \oh \lbr \hat{a} \hat{a}^\da , \hat{\rho} \rbr  \\
\mathcal{\hat{D}}_l^{(2)}[\rho] &=  \hat{a} \hat{a} \hat{\rho} \hat{a}^\da \hat{a}^\da - \oh \lbr \hat{a}^\da \hat{a}^\da  \hat{a} \hat{a} , \hat{\rho} \rbr
\end{align}
Here $\omega_0$ is the cavity frequency and $U/2$ the strength of the Kerr (or Hubbard) interaction.  $\gamma$ is the two-photon loss rate, while $\gamma r$ is the single photon pumping rate.  %This model has recently received attention in the context of quantum synchronization [CITE BRUDER]; it is also directly realizable in superconducting circuit architectures, where strong Kerr interactions and engineered two-photon losses have been experimentally demonstrated [CITE].
We will set $\omega_0=0$ in the following, as it can be eliminated by moving to a rotating frame.
% by the unitary transformation $\hat{U}(t)=e^{i \w_0 t \hat{a}^\da \hat{a}}$,

Note first that the unique steady state density matrix of this model has been previously found analytically in Ref.~\cite{dykman1978,dodonovMizrahiJPAMathGen1997}.  
%A key observation is that the Hamiltonian is diagonal in the photon number basis, whereas the jump terms in the dissipators are strictly off-diagonal in this basis.
The steady state is an incoherent mixture of photon number Fock states; further, it is completely independent of the interaction strength $U$, and is only determined by the  dimensionless parameter $r$ (ratio of the driving to the nonlinear loss).  The photon-number probabilities in the steady state are in fact determined by a classical master equation (i.e.~coherences play no role).
%is completely diagonal in the photon number (Fock) basis and is essentially classical (i.e.~an incoherent mixture of Fock states). Furthermore, the diagonal elements of the steady state density matrix are determined by a classical master equation 
%(no coherences), and only depend on the dimensionless parameter $r$ (the ratio of the driving to the loss) while are completely independent from the local interaction term $U$.
\begin{figure}[t]
\begin{center}
\epsfig{figure=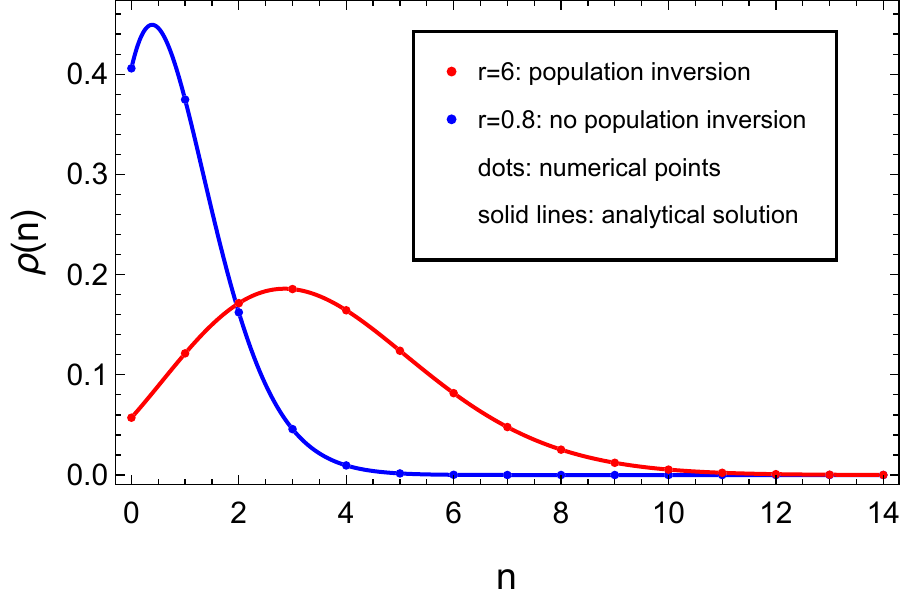,scale=0.85}
\caption{The stationary state density matrix is diagonal in the Fock basis. Its diagonal elements are plotted as a function of the number of bosons for two values of the pump-loss ratio $r$, showing a distribution with and without population inversion.
The stationary state does not depend on the interaction $U$, resonator frequency $\omega_0$ or dissipation scale $\gamma$. Numerical calculations use a Hilbert space cutoff $N_{max}=15$. With this choice of cutoff the numerical solutions (dots) agree perfectly with the analytical predictions (solid lines).}
\label{fig:rhos}
\end{center}
\end{figure}
In Fig.~\ref{fig:rhos} we plot the photon number probabilities $p_n$ in the steady state for two different values of $r$.  For the smaller value of $r$, the probabilities decay monotonically with $n$, whereas for large values, one obtains a peaked, non-monotonic distribution.  As the Hamiltonian $\hat{H}$ dictates that energy increases with increasing photon number, this latter situation corresponds formally to a population inversion.  %The peaked nature of the distribution reflects the simple fact that there is special value of photon number for which the effective loss rate and pumping rates are balanced.
%Still the steady state populations contains some interesting physics as we plot in figure~\ref{fig:rhos}, which shows the stationary state density matrix in the photon number basis, $p_n$ as a function of $r$. We notice that while for small $r$ the weights decay monotonically with $n$, for large values of the ratio between drive and dissipation the stationary state density matrix develops a non-monotonic behavior with $n$ leading to a population inversion, namely higher bosons number states becoming more populated than the corresponding lower number sector. 

A natural question to now ask is whether this inversion effect (which is essentially classical) manifests itself in the cavity's spectral properties. 
Such a question was first addressed in a series of seminal works~\cite{dykman1978,DykmanKrivoglaz84}, where the spectral properties of a related quantum van der Pol oscillator were discussed (see instead Ref.~\onlinecite{dykman1975} for the undriven model). More recently, the power spectrum of a coherently driven quantum van der Pol oscillator was computed to investigate signatures of synchronization~\cite{WalterNunnekampBruderPRL14}.

%is well known since at least twenty years, much less is understood about the excitations on top of the stationary state (though note importantly that a comprehensive discussion of the high temperature regime is given in \cite{dykman1975}). 

% In the next section we will address this question by calculation the single-particle Green function of the model. 

%We start by noting that the unique steady state of this model has been previously found analytically in Ref.~\cite{dodonovMizrahiJPAMathGen1997}.  A key observation is that the Hamiltonian is diagonal in the photon number basis, whereas the jump terms in the dissipators are strictly off-diagonal in this basis.  As a result, the steady state is completely diagonal in the photon number (Fock) basis.  Further, the diagonal elements of the steady state density matrix are determined by a classical master equation (no coherences), and only depend on the dimensionless parameter $r$ (the ratio of the driving to the loss).  

%%%%%%%%%%%%%%%%%%%%%%%%%
\begin{figure*}
\centering
\begin{minipage}[t]{.4\textwidth}
\epsfig{figure=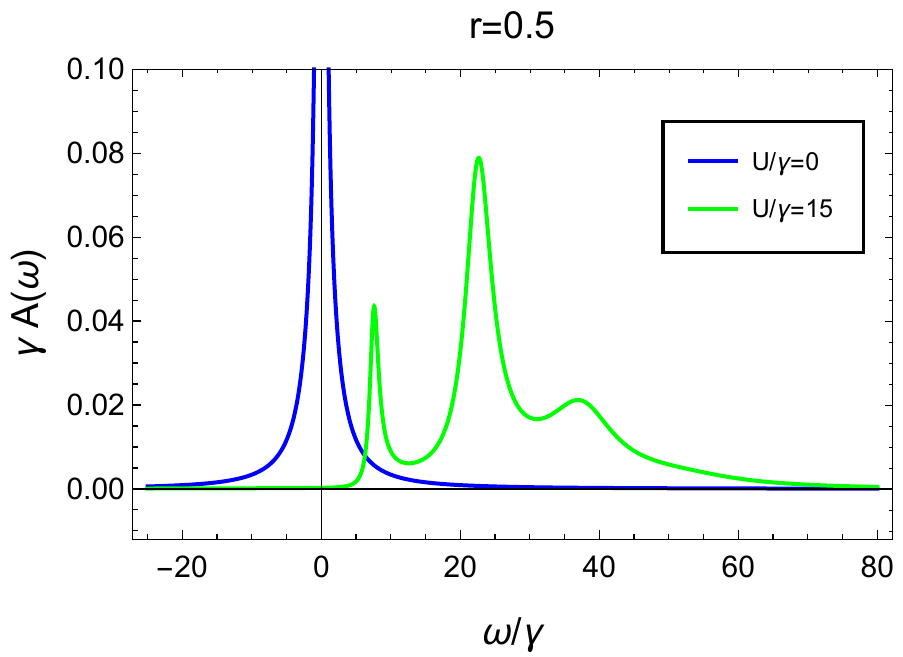,scale=0.8}
\epsfig{figure=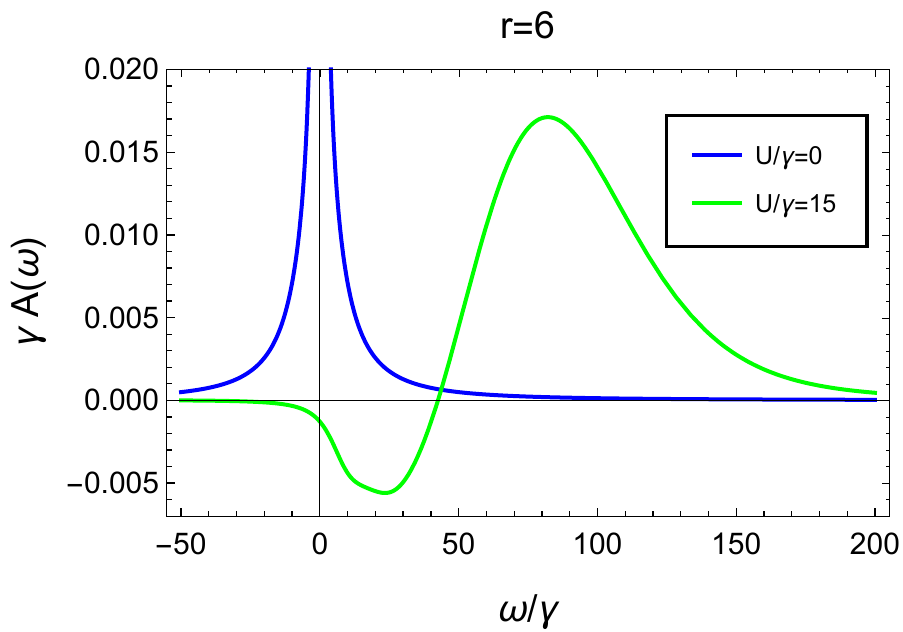,scale=0.8}
\end{minipage}\qquad
\begin{minipage}[t]{.4\textwidth}
\epsfig{figure=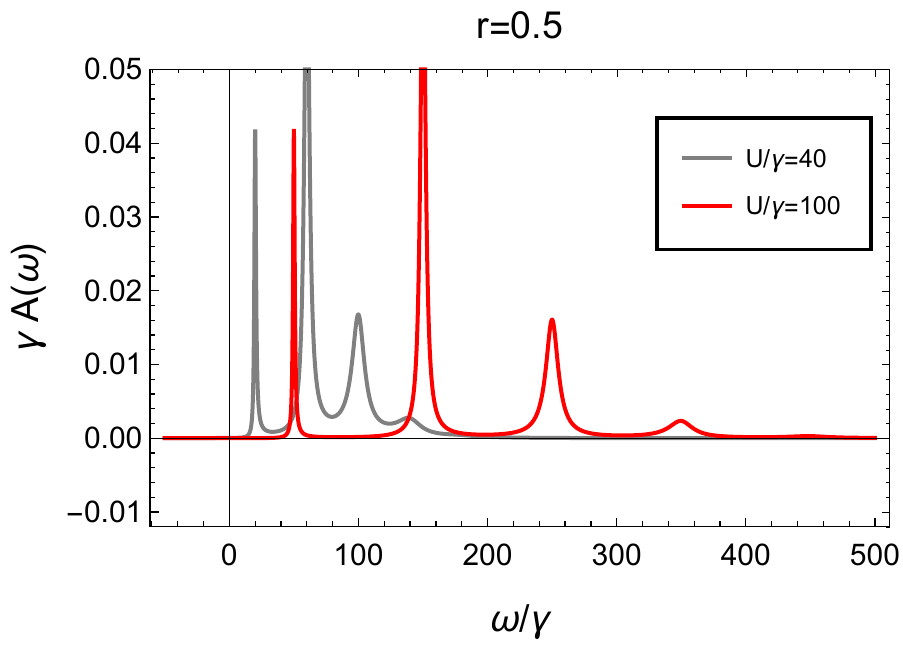,scale=0.8}
\epsfig{figure=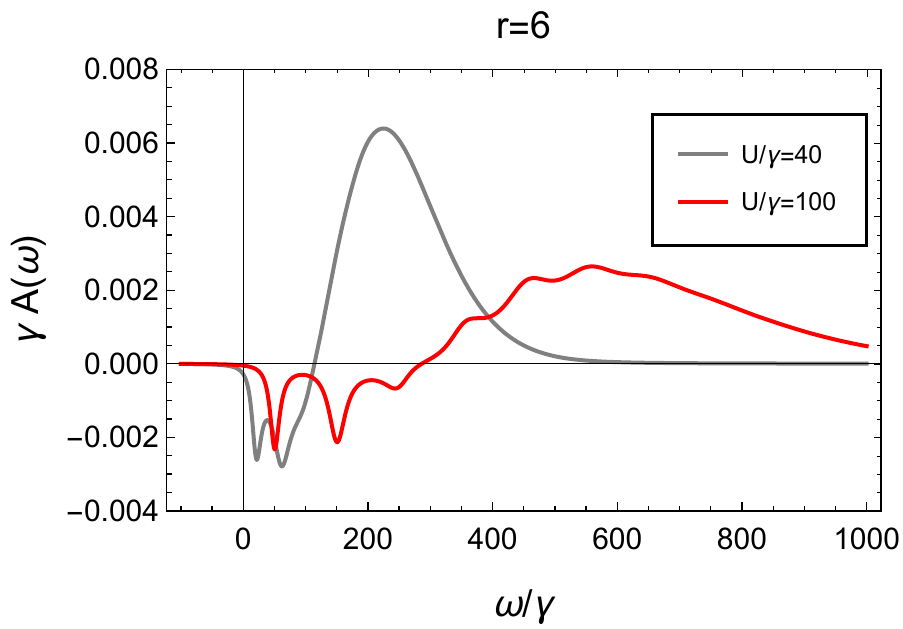,scale=0.8}
\end{minipage}
\caption{Evolution of the single-particle spectral function $A(\w)$ upon changing the interaction $U$, for two values of the parameter $r$, the ratio between drive and losses. For $r=0.5$ (top panel) we see that increasing the nonlinearity splits the single particle peak into a series of well separated resonances. For larger drive, $r=6$ (bottom panel), corresponding to an inverted steady state density matrix, a new feature appear, namely the spectral function becomes negative over a range of frequencies, at least for large enough interaction. Parameters: resonator frequency $\w_0 = 0$, Hilbert space cutoff $N_{max}=15$.
%\AC{Strange to write an explicit value of $\gamma$, what possible difference could it make?  Also, I would just label the y axis $\gamma A(\omega)$, slightly easier to read.}
}\label{fig:retGreens}
\end{figure*}

%%%%%%%%%%%%%%%%%%%%%%%%%%%%%%%%%%%%%%%%%%%%%%%%%%%%%%
\subsection{Liouvillian eigenmodes and symmetry considerations}\label{subsect:sym}

To understand the Green's functions of our model, it will be useful to first discuss its symmetry properties.  Due to driving and dissipation, the system does not conserve photon number. Nonetheless, the Liouvillian is invariant under the $U(1)$ symmetry $\hat{a} \rw \hat{a} e^{i \theta}$.  This implies that the Liouvillian $\lind$ commutes with the superoperator $\hat{\mathcal{K}}= [\hat{a}^\da \hat{a} , \bullet ] $ that generates the symmetry operation. 
As a result, the eigenvalues $k$ of $\hat{\mathcal{K}}$ are quantum numbers which label the eigenstates of $\lind$.   We can use this to write the Liouvillian in the block-diagonal form $\lind= \otimes_k \lind_k $, where $\lind_k$ acts only within the eigensubspace of $\hat{\mathcal{K}}$ corresponding to the (integer) eigenvalue $k$.  We denote the right eigenstates of a particular block $\lind_k$ by
\begin{equation}
	\hat{r}_{\alpha,k} = \sum_n r_{\alpha,k}^{n} \ket{n+k} \bra{n}
\end{equation}
In Fock space, we see that this is a matrix that only has non-zero elements along the $k$th off-diagonal.  

% Here we provide some details on the symmetry of the model as this justifies our perturbative results and simplify numerical computations.
% The model is invariant under the $U(1)$ symmetry $\hat{a}(t)\rw \hat{a}(t) e^{i \theta}$, which is not spoiled by the incoherent drive, but it would have been spoiled by a coherent one, which is more  commonly realized in quantum optics.

% The generator of the symmetry group is the super-operator $\hat{\mathcal{K}}= [\hat{a}^\da \hat{a} , \bullet ] $, which commutes with $\lind$, thus the eigenvalues $k$ of $\hat{\mathcal{K}}$ are quantum numbers which label the eigenstates of $\lind$.
% The eigenstates and eigenvalues of $\lind$ are then labeled by the eigenvalues of $\hat{\mathcal{K}}$ as well, so we use the two indices $\alpha, k$: $\hat{\mathcal{K}}(\hat{r}_{\alpha,k}) = k \hat{r}_{\alpha,k}$.

% In the eigenbasis of $\hat{\mathcal{K}}$, the liouvillian is block diagonal and we can write it as a tensor product of those blocks: $\lind= \otimes_k \lind_k $.
% In the Fock basis, the eigenstates of one block $\lind_k $ have the form $\hat{r}_{\alpha,k} = \sum_n r_{\alpha,k}^{n} \ket{n+k} \bra{n}$, where $k$ measures the distance of the matrix elements of $\hat{r}_{\alpha, k}$ from the principal diagonal.

The presence of this symmetry greatly reduces the numerical complexity of the problem, as we can diagonalize the different blocks separately.  It also gives a simple physical way to label the different eigenmodes of $\lind$.  Eigenmodes corresponding to $k=0$ describe how diagonal elements of the density matrix (in the Fock basis) decay.  Such decay modes conventionally referred to as $T_1$ relaxation processes.  In contrast, eigenmodes corresponding to $k \neq 0$ describe how Fock-state coherences decay.  These are generically referred to as $T_2$ relaxation processes.  

Note crucially that different correlation functions will {\it only} be sensitive to a particular (small) subset of Liouvillian eigenmodes.  
For example, for the single particle Green's function defined in Eq. \eqref{eq:defRetared}, it is only the eigenmodes corresponding to $k=1$ that contribute.  This follows immediately from Eq.~(\ref{eq:lehmannRet}) and the fact that for $k \neq 1$: 
\begin{equation}
	\tr \left( \hat{a} \hat{r}_{\alpha,k} \right) = 0
    \hspace{1 cm}(k \neq 1)
\end{equation}

Analogously, correlation functions like $ G^{(2)} = \aver{ \hat{a}^\da (t) \hat{a}^\da (t) \hat{a}(0) \hat{a}(0) }$ would probe Liouvillian eigenmodes with $k=2$, i.e.~$T_2$ processes involving coherences between states whose photon number differs by $2$.  Similarly, a correlation function destroying $k$ bosons at $t=0$ and creating $k$ bosons at time $t$ would probe the decay of coherences between states whose photon numbers differ by $k$.  It also follows that if one wishes to probe $T_1$ processes (i.e.~$k=0$), one needs to look at density-density correlation functions.

% \old{
% put in a supplementary ? 
% The proof follows from the remark that, if we consider the matrix element $\ket{n+k}\bra{n}$, where $\ket{n}$ is an eigenstate of the Hamiltonian \eqref{eq:hamiltonian}, and we apply the liouvillian to it, we get $\lind \ket{n+k}\bra{n} =  \ket{n'+k}\bra{n'}$; in other words matrix elements with different distance $k$ from the principal diagonal are not coupled by $\lind$.
% This happens generically if all jump operators $L_\mu$ satisfy $[ L_\mu^\da L_\mu, H ] = 0$.

% \begin{figure}[t]
% \begin{center}
% \epsfig{figure=statState,scale=0.7}
% \caption{modulus of eigenstates of the liouv }
% \label{fig:rhos}
% \end{center}
% \end{figure}

%\AC{End of AC edits}

%%%%%%%%%%%%%%%%%%%%%%%%%%%%%%%%%%%%%
\section{Spectral Properties of Driven-Dissipative VDP Oscillator}\label{sectResults}

We now turn to the spectral properties of the nonlinear driven-dissipative cavity model introduced in 
Eq.~\eqref{eq:masterEquation}.
%We will study the properties of the system as a function of the pump/loss ratio $r$ and the interaction $U$, keeping fixed the dissipative scale $\gamma$. 
We use the Lehmann representation given in Eq.~\eqref{eq:lehmannRet} to compute the spectral functions numerically, by truncating the bosonic Hilbert space to a maximum number of states $N_{\rm max}=15$ and diagonalizing $\lind$. 
The cutoff $N_{\rm max}=15$ is enough to obtain accurate results, as it is shown by the agreement of the steady state numerical solution with the analytical prediction in Fig. \ref{fig:rhos}. We further checked that the results for the Green's functions are stable by increasing $N_{\rm max}$ and that they satisfy sum properties like Eq. \eqref{eq:sumProp}.

\subsection{Spectral function and the role of interactions}

In Fig.~\ref{fig:retGreens} we plot the spectral function $A(\omega)$ (c.f.~Eq.~\eqref{eq:specFunc}) of our system for several values of the dimensionless interaction strength $U/\gamma$ and for two values of the drive/loss ratio $r$.  An immediate result, visible in all four panels, is that {\it the spectral functions strongly depend on the interaction strength}.  This dependence is remarkably different from the steady state density matrix, which (as discussed in Sec.~\ref{sectModel}) is completely insensitive to $U$.  Heuristically, while the steady state density matrix is completely independent of the system's coherent Hamiltonian dynamics, the system's response to perturbations retains a strong dependence on $\hat{H}$.  
%A first interesting result that clearly emerges is that the steady state spectral properties of the model strongly depends on $U$, while the steady state density matrix does not, as we discussed in section~\ref{sectModel}. 
%In other words, excitations created on top of the stationary state retain a strong dependence from the coherent part of the Lindbladian.
%low drive regime and Hubbard Bands

For a more detailed analysis, consider first the regime of relatively weak driving where $r=0.5$ (top row of Fig.~\ref{fig:retGreens}).  To understand lineshapes, recall from Sec.~\ref{subsect:sym}  that the spectral function is probing $T_2$ decay modes which describe the decay of coherences between Fock states $|n\rangle$ and $|n + 1\rangle$.  The oscillation frequency of these coherences is largely determined by the coherent Hamiltonian $\hat{H}$.  For $U=0$, there is no Hamiltonian, and coherences do not oscillate; we thus obtain a single peak in the spectral function.  As $U/\gamma$ is increased, distinct peaks become visible in $A(\omega)$ (each approximately Lorentzian), corresponding to different coherences and different decay modes; the peaks become more and more resolved with increasing $U/\gamma$.  Note that in this weak driving regime, there is no obvious signature of non-equilibrium in the spectral function.

\begin{figure}[t]
\begin{center}
\epsfig{figure=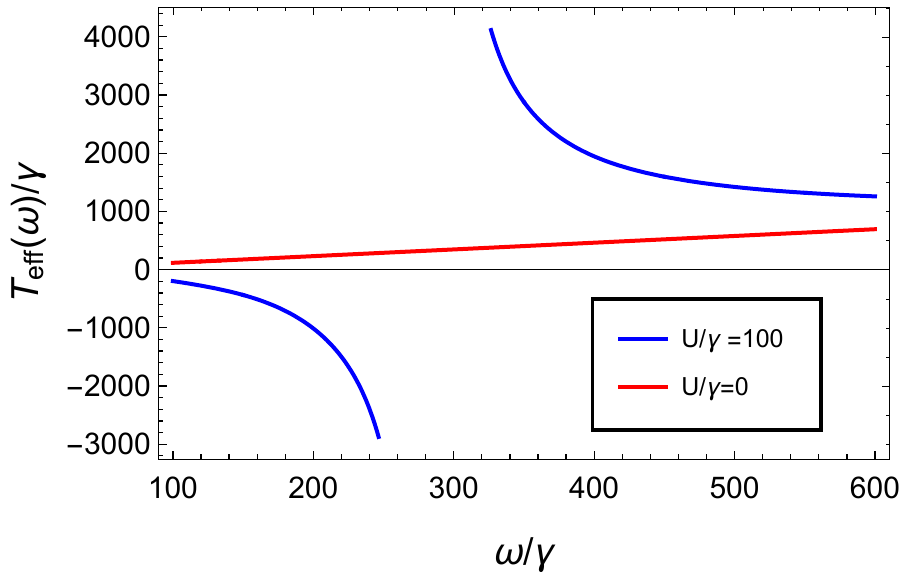,scale=0.8}
\caption{Frequency-dependent effective temperature 
$T_{\text{eff}}(\w)$ as defined in Eq.~(\ref{eq:effTemp}) in the regime of large pump-loss ratio ($r=6)$. We notice that  interaction $U$ makes the effective temperature  turn negative in some \emph{positive} frequency region. For comparison we also plot the effective temperature of a $U=0$ cavity, which is always positive for $\omega>0$. Parameters: resonator frequency $\w_0 = 0$, Hilbert space cutoff $N_{max}=15$.}
\label{fig:effTemp}
\end{center}
\end{figure}

For larger values of the driving parameter $r$ (bottom row of Fig.~\ref{fig:retGreens}, $r=6$), the situation is markedly different.  For large driving and large enough interaction $U$, we find that the spectral function hits zero at a positive finite frequency, and for larger frequencies, becomes negative.  We term this negativity of $A(\omega)$ at $\omega > 0$ a ``negative density of states" (NDoS).  This is a clear indicator of non-equilibrium: as discussed in Sec.~(\ref{subsect:GFclosed}), this cannot happen in a system that is in thermal equilibrium.  We also stress that (as discussed in Sec.~\ref{subsec:effTemp} and in \cite{lavitanClerkNJP2016}), this NDoS corresponds to a negative effective temperature; this is shown in Fig. \ref{fig:effTemp}.  As also discussed, this negative temperature effect could be directly probed by coupling the cavity weakly to an auxiliary probe qubit.  
%
%\textcolor{green}{\textit{I think the name "negative density of states" is not appropriate because the dos is negative in equilibrium for neg frequencies as well.}}
%\AC{No issue here, clear that we are talking about the behaviour at positive frequencies.}

%We stress that this cannot happen in equilibrium, where single particle spectral functions have the sign property of Eq. \eqref{eq:signPropA}, implying that they must change sign only at $\w=0$, but such a constraint does not hold in general for a non-equilibrium stationary state. 

One might first think that the NDoS effect here is simply a reflection of the population inversion in the steady state photon number distribution, which occurs when $r$ is sufficiently large.  This is not the case:  while the population inversion in the steady state is independent of $U/\gamma$, $A(\omega)$ only becomes negative at $\omega > 0$ for sufficiently large $U/\gamma$.  This is shown explicitly in Fig.~\ref{fig:retGreens}.  The relation between the NDoS effect in the spectral function and population inversion in the steady state is thus not entirely trivial; we will explore this in more detail in the next sections. Note that similar spectral function negativity in presence of a population inversion has previously been identified in a related model of a quantum van der Pol oscillator in presence of negative damping and monochromatic drive~\cite{dykman1978,DykmanKrivoglaz84}, as well as in parametrically driven bosonic systems \cite{lavitanClerkNJP2016}.

\subsection{Dissipation-Induced Lifetime}\label{subsect:lifetime}

%\AC{Motivate this section: does dissipation do more than simply broaden the resonances we would expect from knowing the steady state?  Is the broadening describable from simple FGR calculations?}
% \comm{I can apply non-deg pert theory even if the system is degenerate thanks to the symmetry of this specific model, as I only need to compute corrections in the non-diagonal blocks of the liouvillian, which are non degenerate. This perturbation theory treatment to compute single particle excitation lifetimes cannot be generalized, so I decided not to promote it to the previous section}

The results of the previous section show that the spectral properties of the nonlinear quantum VdP oscillator are remarkably rich.  In this section, we investigate the extent to which these can be understood using a perturbative approach where the only dynamical effect of dissipation and driving taken into account is to give a finite lifetime to the Fock-state eigenstates of the system Hamiltonian $H$.  

%%%%%%%%%%%%%%
%The results of previous section have revealed a rich structure in the spectral properties of our quantum VdP oscillator which turns out to be  strongly dependent from interaction and non-equilibrium effects. It also have left open the question of the relation between these features and the information contained in the stationary density matrix. To gain some further insight, in this section we ask whether it is possible to capture some of the spectral features of the problem within a simpler approach that starts from the knowledge of the steady state density matrix  and builds a spectral function assuming the system to be only weakly open. In other words, we assume that the main effect of dissipation is (i) to establish to a non-trivial stationary state with population $p_n$ and (ii) to broaden with a finite lifetime the transitions for going to state $n$ to state $n+1$.

%In this section, we analyze how qualitatively different behaviors of the spectral function arise from Eq. \eqref{eq:lehmannRet}, understanding the role of dissipation in this respect.
%The first noticeable and most expected dissipative effect is that one particle excitations acquire a finite decay rate $\Gamma_\alpha =-\re{\lambda_\alpha}$. \new{One could be tempted to think that this is the only effect of dissipation of Green's functions, so here we take this path} and
%We approximate the spectral function of the open system just by adding this inverse lifetime to the equilibrium Lehmann's decomposition of the spectral function of Eq. \eqref{eq:lehRetEq}. 
%%%%%%%%%%%%%%%

Our starting point is the open-system Lehmann representation of Eq. \eqref{eq:lehmannRet}.  We will approximate the eigenstates of the Liouvillian to be the same as those of the closed system, e.g. simple outer products of Fock states (c.f. Eq.~\ref{eq:eigstatZeroOrd}), ignoring their perturbative corrections.  We will however take into account the modification of the Liouvillian's eigenvalues to leading order in the dissipation (i.e.~in $\gamma$).  Formally, this procedure can be implemented using the Lindblad perturbation theory approach introduced in Ref.~\cite{liPetruccione2014SciRep2014}.  
We write our full Liouvillian as  $\lind = \lind^{(0)} + \hat{\mathcal{D}}$, 
with the unperturbed Liouvillian $\lind^{(0)}[\rho] = - i [\hat{H},\hat{\rho}] $ and the perturbation $\hat{\mathcal{D}} = \gamma \lp r \hat{\mathcal{D}}_p^{(1)} +  \hat{\mathcal{D}}_l^{(2)} \rp$.
%and we expand around the closed system, considering the dissipator $\hat{\mathcal{D}} = \gamma \lp r \hat{\mathcal{D}}_p^{(1)} +  \hat{\mathcal{D}}_l^{(2)} \rp$ as the perturbation. 
While $\lind^{(0)} $ is highly degenerate, one can still employ the simple non-degenerate perturbation theory of Ref.~\cite{liPetruccione2014SciRep2014} due to the symmetry of $\lind$ discussed in Sec.~\ref{subsect:sym}.  This symmetry prevents mixing between different degenerate sectors.

%We must say that the unperturbed liouvillian $\lind^{(0)} $ is highly degenerate, as each eigenstate of $\hat{H}$ corresponds to an eigenvector of $\lind^{(0)}$ with zero eigenvalue, but it is still safe to apply the non-degenerate perturbation theory discussed in \cite{liPetruccione2014SciRep2014} for the specific model considered, as, thanks to the symmetry of the liouvillian $\lind$, it turns out that we only need to compute the corrections to eigenvalues corresponding to non-degenerate sectors.
We expand eigenvalues and eigenvectors of the Liouvillian in powers of $\gamma$: \begin{eqnarray*}
\lambda_\alpha = \lambda_\alpha^{(0) }+ \lambda_\alpha^{(1) } + O(\gamma^2) \\ 
\hat{r}_\alpha = \hat{r}_\alpha^{(0)} + \hat{r}_\alpha^{(1)}+ O(\gamma^2) \\ \hat{ l}_\alpha = \hat{l}_\alpha^{(0)} + \hat{l}_\alpha^{(1)} + O(\gamma^2)
\end{eqnarray*}
with the unperturbed quantities $\lambda^{(0)}_\alpha, \hat{l}^{(0)}_\alpha, \hat{r}^{(0)}_\alpha$ already defined in Eqs.~\eqref{eq:eigvalZeroOrd},\eqref{eq:eigstatZeroOrd}.  
We retain the perturbative corrections to the eigenvalues, while ignoring for the time being any corrections to the eigenstates. The validity of such an approximation and the role of these corrections will be discussed later in the manuscript. Perturbation theory tells us that the leading order correction to the Liouvillian eigenvalues $\lambda_\alpha$ are given by  $\lambda^{(1)}_{\alpha} =  \tr{ \lsq ({\hat{l}^{(0)}_{\alpha}})^\da \hat{ \mathcal{D}} \, ( \hat{r}^{(0)}_{\alpha} ) \rsq}$. Accordingly,  Eq. \eqref{eq:lehmannRet} yields the following approximate form for the spectral function:
%From the Lehmann representation of Eq. \eqref{eq:lehmannRet}, keeping the eigenstates $\hat{r}_\alpha^{(0)},\hat{l}_\alpha^{(0)} $ at zero order and using the fact that $\hat{\rho}_s$ is diagonal, we obtain
%\beq
%G^R (\w) = \sum_{i,j} \frac{ \bra{\psi_j} a \ket{\psi_i} }{\w - (E_i -E_j) + i \lind^{(1)}_{i,j}  } \lp  \bra{\psi_j} a^\da \rho_s \ket{\psi_i} - \bra{\psi_j} \rho_s a^\da \ket{\psi_i} \rp 
%\eeq
%
%In our case, $\rho_s$ is diagonal and $\ket{\psi_i} = \ket{n}$ thus 
%\beq
%\label{eq:1stOrdLeh}
%G^R (\w) = \sum_{n=0}^\infty \frac{ \abs{\bra{n+1} a^\da \ket{n} }^2}{\w - \Delta_{n+1,n}+ i \Gamma_{n+1,n} }\lp  p_n  - p_{n+1}\rp 
%\eeq
\beq
\label{eq:1stOrdLeh}
A(\w) =\frac{1}{\pi} \sum_{n=0}^\infty  \frac{ \Gamma_{n+1,n} \abs{\bra{n+1} \hat{a}^\da \ket{n} }^2 \lp  p_n  - p_{n+1}\rp }{\lp \w - E_{n+1,n} \rp^2 +  \Gamma_{n+1,n}^2 }
\eeq
where
\begin{eqnarray}
\label{eq:1stOrdExcitEn}
E_{n+1,n} & = & - \im \left( \lambda_{n+1,n}^{(0)} + \lambda_{n+1,n}^{(1)} \right) \nonumber \\
            & = & E_{n+1} - E_n = \w_0  + U/2 + U n \\
\label{eq:1stOrdLifetime}
\Gamma_{n+1,n}  & = & - \re \left( \lambda_{n+1,n}^{(0)} + \lambda^{(1)}_{n+1,n} \right) \nonumber \\
        & = & 2 \gamma n^2 + r \gamma  (2n+3)
\end{eqnarray}
Note that the first order correction to the $\lambda_{\alpha}$ is purely real, implying there is no shift in the position of the spectral function resonances.  The approximate spectral funciton in Eq.~\eqref{eq:1stOrdLeh} is exactly the same as the equilibrium expression in Eq.~\eqref{eq:lehRetEq1}, except that the populations $p_n$ are non-thermal, and each resonance has a finite width $\Gamma_{n+1,n}$.  While we have shown how this width can be calculated using formal perturbation theory, it also has a simple physical origin: it is the sum of the Fermi's Golden rule decay rates for the states $|n\rangle$ and $|n+1\rangle$, i.e. 
\begin{eqnarray}
\es{
\Gamma_{n+1, n} = \sum_m & r \gamma \lp  \abs{ \bra{m} \hat{a}^\da \ket{n} }^2 + \abs{ \bra{m} \hat{a}^\da \ket{n+1} }^2 \rp +  \\
+ \sum_m  &\gamma \lp \abs{ \bra{m} \hat{a}\hat{a} \ket{n} }^2  +\abs{ \bra{m} \hat{a}\hat{a} \ket{n+1} }^2  \rp
}
\end{eqnarray}
The first line is the decay rate due to the incoherent driving, the second due to the two-photon loss.

%%%%%%%%%%%%%%%%%%%%%%
\begin{figure}[t]
\begin{center}
\epsfig{figure=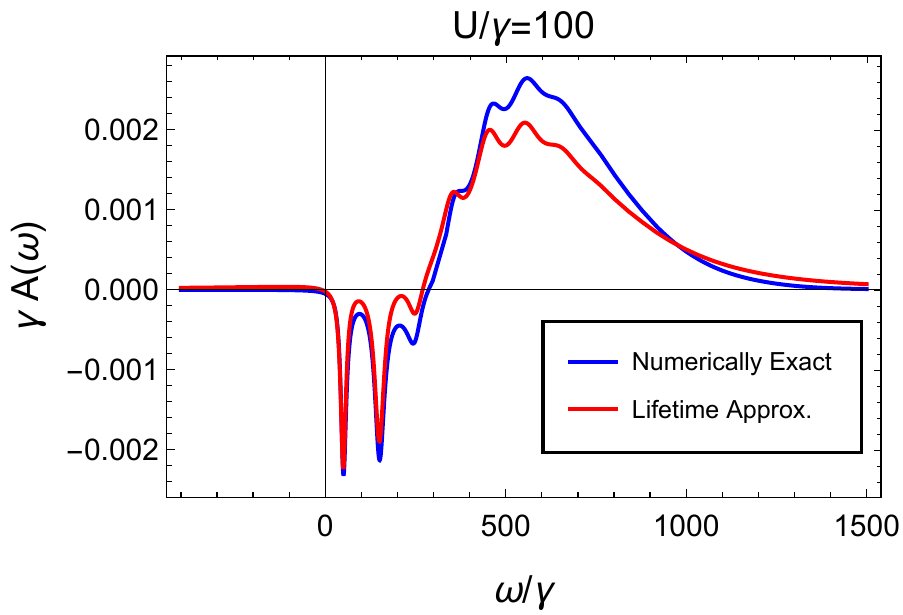,scale=0.8}
\epsfig{figure=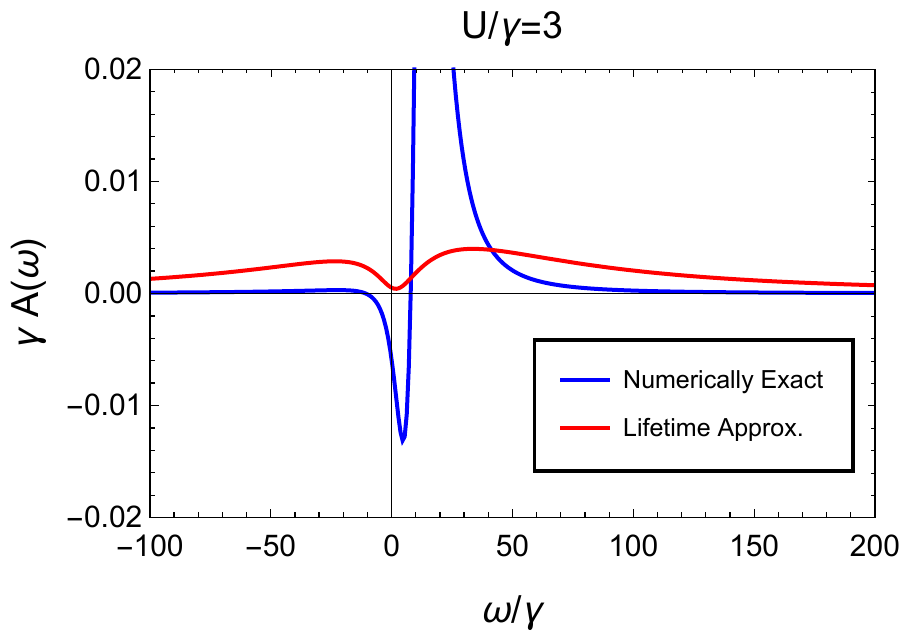,scale=0.8}
\caption{The spectral function $A (\w ) $ obtained by numerical evaluation of the exact Lehmann representation, as well as that obtained using the lifetime approximation of Eq. \eqref{eq:1stOrdLeh}.
Top: The perturbative treatment of the lifetime is in good agreement with the exact result for a value of the interaction $U/\gamma=100$.  Bottom: Perturbation theory gets bad when the resonances are not well resolved. Here the interaction is $U/\gamma = 3$.
Parameters: resonator frequency $\w_0 = 0$, pump-loss ratio $r=6$, Hilbert space cutoff $N_{max}=15$.
%\AC{First line in legend should be ``Numerically exact".  Also, again, no need to write value of $\gamma$. }
}
\label{fig:1stOrdRet}
\end{center}
\end{figure}
%%%%%%%%%%%%%%%%%%%%%%

%Populations are in fact set by the ratio $r$ between drive and dissipation, resulting in the non-equilibrium distribution computed in \cite{dodonovMizrahiJPAMathGen1997}.
%In particular, for a big enough value of the drive (the vertical line in Fig. \ref{fig:mapZeroSpecFunc}), the stationary state undergoes population inversion, as we show in Fig. \ref{fig:rhos}, which we defined as a violation of the condition of Eq. \eqref{eq:nonPopInv}.

In Fig.~\ref{fig:1stOrdRet} we compare the perturbative result with the full calculation obtained with the Lehmann representation for $r=6$ and two values of the Kerr interaction. We see that at large $U/\gamma\simeq 100$ the perturbative approach captures rather well the main features of the spectrum, in particular the location of the peaks, their width and weight. However, upon decreasing the interaction, the agreement deteriorates, as we show for $U/\gamma=3$.
This behaviour is of course not surprising, as the perturbative approach is only valid in the small dissipation limit $1 \ll U/\gamma$,$r  \ll U/\gamma$.  
% In fact, our perturbative treatment of the dissipator is valid for small dissipation with respect to interaction, that is for $\gamma /U,r \gamma /U << 1 $.
As a rough rule of thumb, when resonances in Eq.~\eqref{eq:1stOrdLeh} begin to overlap, perturbation theory starts getting bad, as the spacing between adjacent resonances is of order $U$ and the width of the resonances of order $\gamma$.

Taking into account the dissipation-induced lifetime in Eq.~\eqref{eq:1stOrdLeh} allows to uncover a mechanism by which dissipation can mask the effect of a population-inverted density matrix on the spectral function. 
Indeed, a population inversion in the stationary state, if there were no lifetime broadening, would certainly result in a violation of the Green's functions sign property in Eq.~\eqref{eq:signPropA}, as one can see straight from Eq.~\eqref{eq:lehRetEq1}. 
%A new sign rule would establish: for $\w>0$, for example, $A(\w)$ would be negative up to $\w_M=E_{M+1}-E_M$, with $E_M$ corresponding to the highest population $p_M$, and positive afterwards. 
On the other hand, the lifetime broadens the resonances, making them overlap and possibly resulting in those with smaller weights to be completely masked by bigger ones. As a result, the spectral function in Eq.~\eqref{eq:1stOrdLeh} does not obey anymore a precise sign rule which is dictated by the behaviour of populations of the density matrix.
As a corollary, the presence of population inversion in the stationary state may not be revealed by a change of sign of the spectral function.
%In our model we distinguish two regimes, which are evident in Fig. \ref{fig:retGreens}: for $U/\gamma >> 1$ resonances are well resolved and an inversion of populations is directly reflected in a negative region for $A(\w)$, while for $U/\gamma \lesssim 1$ resonances overlap washing away this correspondence and $A(\w)$ is always positive. 
%This behavior is expected, as at first order in perturbation theory, the distance between two lorentzians increases linearly with $U$ (Eq.~\eqref{eq:1stOrdExcitEn}), while the broadening is of order $\gamma$ (Eq.~\eqref{eq:1stOrdLifetime}).
 %Ultimately, for vanishing $U$, resonances merge into a single broad one, akin to the spectral function of the damped harmonic oscillator.
% \AC{This paragraph could be massively shortened, as the point here is simple.  If all the resonances were distinct, population inversion in the steady state gives you Lorentzians with negative weights, hence negative DOS.  However, with lifetime broadening, the resonances overlap.  Hence, positive resonances can mask the negativity you would get from negative resonances.  Should just say this in 2-3 sentences.}

%\new{We remark that while changing $U$, population inversion is always present in the stationary state, but a threshold $U$ is needed in order to see its effect in the spectral function.}

\begin{figure}[t]
\begin{center}
\epsfig{figure=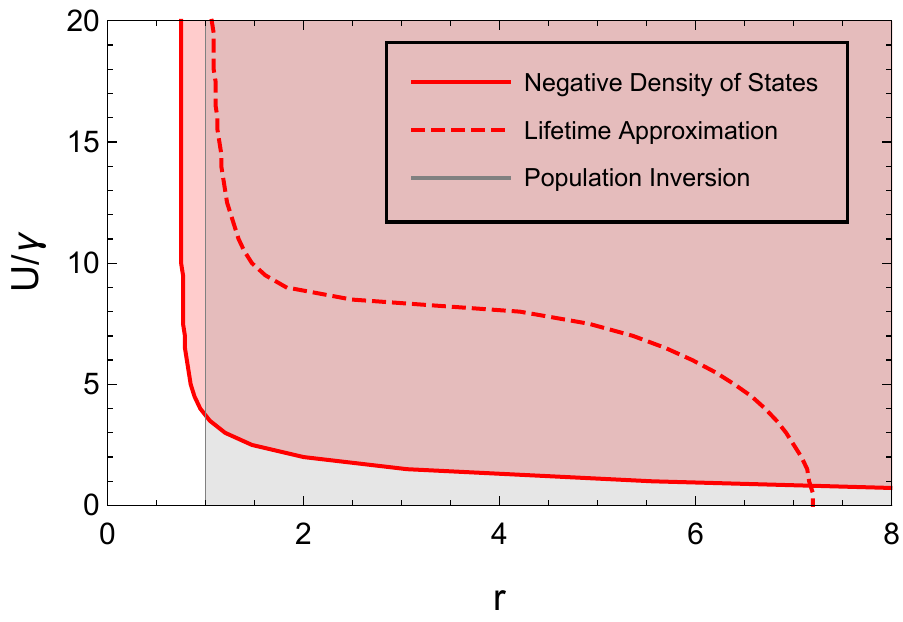,scale=0.8}
\caption{Region of parameters in the $r$-$U$ plane  where the spectral function $A(\w)$ is negative in some positive frequency range (NDoS), according to the exact result obtained by the Lehmann representation, Eq.~(\ref{eq:lehmannRet}), and to the lifetime approximation of Eq. \eqref{eq:1stOrdLeh}. The vertical grey line shows the threshold value of $r$ for which population inversion sets in the stationary state. According to the lifetime approximation, a NDoS is only possible with population inversion in the stationary state, while the exact result shows that this is not strictly necessary. Parameters: resonator frequency $\w_0 = 0$, Hilbert space cutoff $N_{max}=15$.}
\label{fig:mapZeroSpecFunc}
\end{center}
\end{figure}

In Fig.~\ref{fig:mapZeroSpecFunc}, we summarize the above analysis by presenting a phase diagram, in the $(r,U/\gamma)$ plane, the regions of parameter space exhibiting the NDoS effect (i.e.~spectral function $A(\omega)$ is negative at positive frequencies).  The region $r > 1$ (shaded grey) indicates where the steady-state exhibits a population inversion; this boundary can be determined analytically from the exact stationary state solution \cite{dodonovMizrahiJPAMathGen1997} and we remark that it is independent of $U$. In contrast, the spectral function is sensitive to both interaction and non equilibrium effects, resulting in a non-trivial value $U_c(r)$ above which the negative density of state emerges.  We plot this threshold interaction strength both for the numerically exact calculation of the spectral function (red-solid line), and for the approximate perturbative (lifetime broadening) calculation (red-dashed line).  In general, the perturbative approach underestimates the NDoS effect; further, it fails to yield any population inversion in the region $r < 1$.  In contrast, the numerically exact calculation reveals that NDoS can occur even for $r < 1$, i.e.~in regions where the steady state photon number exhibits no population inversion.  This is a remarkable result, which points toward yet another origin of NDoS, as we are going to further discuss below.

\subsection{Dissipative effects beyond lifetime broadening}

As demonstrated above, the simple (perturbative) lifetime broadening of eigenstates introduced in Eq.~(\ref{eq:1stOrdLeh})  was able to capture many aspects of the spectral function of our model.  It however failed to describe the most interesting aspect of Fig.~\ref{fig:mapZeroSpecFunc}:  there are parameter regions where the spectral function exhibits NDoS, even though the steady state density matrix does not exhibit population inversion.  As we now show, this effect can also be captured in perturbation theory if we go beyond simply calculating a correction to the Liouvillian eigenvalues due to dissipation, but also calculate the change to the eigenmodes themselves.  The leading eigenmode correction can cause the weight factors $w_{\alpha}$ in Eqs.~(\ref{eq:lehmannRet}) and to acquire an imaginary part, implying that the spectral function is no longer a simple sum of Lorentzians.  This provides a new route for NDoS.

Using the same perturbation theory used in Sec.~\ref{subsect:lifetime}, we can analytically compute the leading-order-in-$\gamma$ correction to the Liouvillian eigenmodes.  
Following \cite{liPetruccione2014SciRep2014} and using the existence of $\lind^\mo$, the first order corrections to the right and left eigenstates are given by:
%(as it is the case here, in order to obtain the 1st order corrections to left and right eigenstates of $\lind^{(0)}$, we compute the corrections to the right eigenstates of $\lind^{(0)}$ and ${(\lind^{(0)})^\da}$. Then we obtain
\begin{eqnarray}
\hat{r}_{\alpha}^{(1)} = \sum_{\beta \neq \alpha} \frac{\tr \lsq { (\hat{r}_\beta^{(0)}) }^\da \hat{\mathcal{D}} ( \hat{r}_\alpha^{(0)} ) \rsq}{\lambda_\alpha^{(0)}  - \lambda_\beta^{(0)}  } \hat{r}_\beta^{(0)} \\
\hat{l}_{\alpha}^{(1)} = \sum_{\beta \neq \alpha} \frac{\tr \lsq { (\hat{l}_\beta^{(0)}) }^\da \hat{\mathcal{D}}^\da (\hat{l}_\alpha^{(0)} ) \rsq}{{\lambda_\alpha^{(0)}} ^*  - {\lambda_\beta^{(0)} }^* } \hat{l}_\beta^{(0)} 
\end{eqnarray}
As expected, dissipation mixes the various eigenmodes together with a strength that is inversely proportional to the difference in eigenvalues.  Here, the denominator is purely imaginary (as all unperturbed eigenvalues are imaginary).  
%This result resembles the standard quantum mechanical perturbative correction to eigenstates, proportional to the matrix element of the perturbation (in this case the dissipator) divided by the energy difference which in this case is purely imaginary since we perturb around the closed system limit. 

As discussed, for the spectral function, the unperturbed modes of interest correspond to coherences between the $\ket{n}$ and $\ket{n+1}$ Fock states:
\begin{eqnarray}
    \hat{r}^{(0)}_{n+1,n} & = \hat{l}^{(0)}_{n+1,n} = \ket{n+1}\bra{n}
\end{eqnarray}
With dissipation, these modes acquire a real part to their eigenvalues, corresponding to dephasing.  The first order correction to the mode wavefunctions take the form:
\begin{eqnarray}
\label{eq:1stOrdStatesR}
\es{ 
\hat{r}_{n+1,n}^{(1)} =&- i \frac{r\gamma}{U} 2 \sqrt{(n+2)(n+1)} \, \hat{r}_{n+2,n+1}^{(0)} + \\ &+ i \frac{\gamma}{U} n \sqrt{n^2-1} \, \hat{r}_{n-1,n-2}^{(0)}} \\
\label{eq:1stOrdStatesL}
\es{
\hat{l}_{n+1,n}^{(1)} =&- i \frac{r \gamma}{U} 2 \sqrt{(n+1)n} \, \hat{l}_{n,n-1}^{(0)} + \\ &+ i \frac{\gamma}{U} (n+2) \sqrt{(n+1)(n+3)} \, \hat{l}_{n+3,n+2}^{(0)}}
\end{eqnarray}
%where $\hat{r}^{(0)}_{n+1,n} =\hat{l}^{(0)}_{n+1,n} = \ket{n+1}\bra{n}$ are the unperturbed eigenstates.
At a physical level, these corrections tell us that dephasing eigemodes of the Liouvillian no longer correspond to a single Fock state coherence; rather, each mode involves three distinct coherences.

%%%%%%%%%%%%%%%%%%%
\begin{figure}[t]
\begin{center}
\epsfig{figure=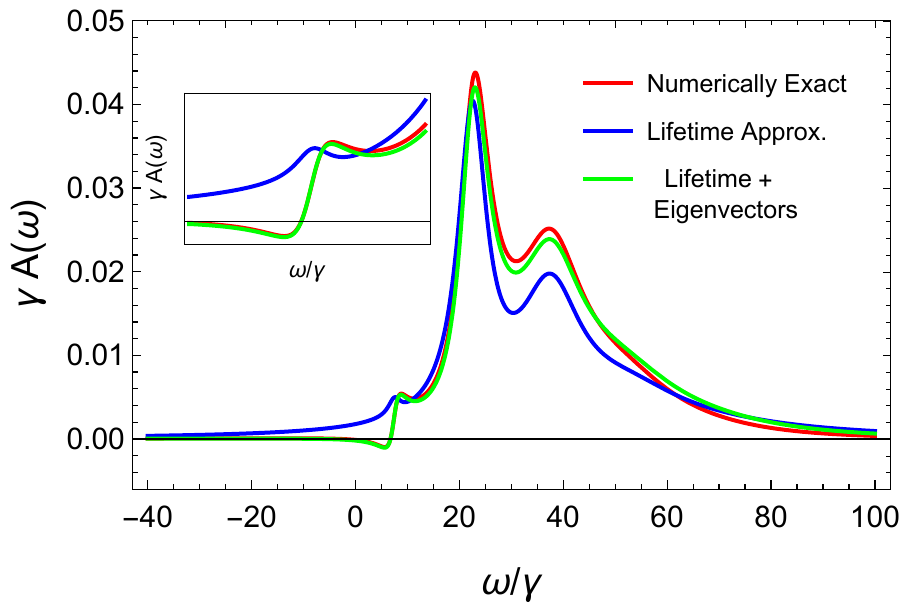,scale=0.8}
\caption{The spectral function $A(\w)$ for a value of $r$ just below the threshold needed in order to have steady state population inversion. Strikingly, the spectral function $A(\w)$ (as computed numerically) still exhibits negativity at positive frequencies.  This feature is missed if one calculates $A(\w)$ using the simple lifetime approximation of Eq. \eqref{eq:1stOrdLeh}. Including the dissipative correction to the Liouvillian eigenstates (to leading order), one is then able to recover the negative part of $A(\w)$. 
Parameters: resonator frequency $\w_0 =0$, interaction $U/\gamma=15$ , pump-loss rate $r=0.94$, Hilbert space cutoff $N_{max}=15$.}
\label{fig:specBeyoLT}
\end{center}
\end{figure}
%%%%%%%%%%%%%%%%%%%%

 In Fig.~\ref{fig:specBeyoLT} we show the effect of including these eigenmode corrections in the evaluation of the spectral function. We see that this modified approach is able to capture non-Lorentzian contributions to the spectral function, and to improve qualitatively and quantitatively the agreement with the exact numerical result. In particular, a region of negative density of states now appears at small frequency, an effect which is completely missed by the lifetime broadening approximation.
 
These eigenmode corrections can also be given a physical interpretation in terms of interference of different dephasing modes.  
%Note first that as the steady state density matrix is diagonal in the Fock basis, it corresponds to a statistical uncertainty as to the photon number.  
Consider the contributions to the time-domain correlation function in Eq.~\eqref{eq:lehmannGreat} associated with a particular initial photon number $m$:  
\beq 
\label{eq:weightGreat}
\sum_n e^{\lambda_{n+1,n}t }
\lp \sum_l \bra{l} \hat{a} \hat{r}_{n+1,n} \ket{l}  \rp \bra{m} \hat{l}_{n+1,n}^\da \hat{a} ^\da \ket{m} \hat{\rho}_{m,m} \eeq
Recall the interpretation: starting with $m$ photons, we add a photon to the cavity, exciting a dephasing eigenmode $\alpha = (n+1,n)$ of the Liouvillian.  To $0$-th order in dissipation, the time-independent weight factors are necessarily real.  This follows from the fact that i) $\hat{r}^{(0)}_{n+1,n} =\hat{l}^{(0)}_{n+1,n}$, and
%, as the hamiltonian is hermitian 
ii) the only non-zero contribution is when $l=n=m$, i.e.~adding a photon to $\ket{m}$ excites a single, unique dephasing eigenmode.
%$\hat{r}^{(0)}_{n+1,n}$ is has a single non-zero matrix element in the basis in which $\rho_s$ is diagonal.

Including dissipation to first order, both conditions (i) and (ii) no longer hold.  In particular, as the dephasing eigenmodes no longer correspond to a single Fock coherence (c.f.~Eq.~(\ref{eq:1stOrdStatesR})-(\ref{eq:1stOrdStatesL})), adding a photon to $\ket{m}$ can simultaneously excite several distinct dephasing eigenmodes.  It is the interference between these processes that give rise to complex weights and hence non-Lorentzian contributions to the spectral functions.  The spectral function is thus sensitive to an interference in the dynamics, even though there is no coherence in the steady state density matrix.

%In presence of a dissipator, condition i) doesn't hold anymore as the liouvillian is not hermitian nor condition ii) does, as a single eigenstate of the liouvillian couples several coherences (and populations, in the general case), which "interfere" in Eq. \eqref{eq:weightGreat}.
%As a result, already introducing 1st order corrections Eqs.~\eqref{eq:1stOrdStatesR},\eqref{eq:1stOrdStatesL}, the weights of single particle Green's functions turn complex, introducing non-lorentzian contributions in the spectral function. 

Stepping back, we thus see that even for weak dissipation, the spectral function is sensitive to more than just the lifetime-broadening effect of dissipation:  the fact that dissipation can also create more complicated dephasing processes also directly impacts the form of $A(\omega)$.  This gives rise to anti-Lorentzian contributions, and (in our model) negative density of states in regimes where the steady state exhibits no population inversion.
%To conclude this section, we have shown that broadening of resonances is not the only effect of dissipation on correlation functions and that indeed dissipation changes the functional form of the spectral function, introducing anti-lorentzian contributions. This leads, in our model, to a negative density of states even in the absence of population inversion in the density matrix. 

%\AC{The results in this section are very interesting, but a bit formal.  Is there some way to describe in physical terms what the corrections in Eq. 29 and 30 represent, and why this might induce population inversion?  Heuristically, anti-Lorentzians usually correspond to energy shifts; somehow these are now entering in what we would normally consider a DOS?}

%Interpretation: What is the meaning of the negative DOS here???? Negative Effective Temperature (Freq dependent???)

\section{Conclusions}\label{sectConcl}

In this work we have studied the spectral properties of driven-dissipative quantum systems,
taking the simple case of a quantum van der Pol oscillator as a working example. We have first derived some general results concerning the single particle Green's function of systems described by a Lindblad Master Equation. Using a decomposition in terms of exact eigenstates of the Liouvillian we have derived a Lehmann like representation for the Green's function and compared it to the well known result for closed systems in thermal equilibrium.
Such a result, in addition of being of practical relevance for numerical computations whenever the system is sufficiently small to be diagonalized exactly,  has also a conceptual value. 
From one side it connects properties of the Liouville eigenvalues and eigenstates, which are of theoretical interest but often hard to access, to the behavior of the spectral functions, which are of direct experimental relevance. In addition it allows for a more transparent interpretation of spectral features in regimes far from equilibrium, for which a simple intuition is often lacking or misleading. As an example we have shown that the well known sign property of equilibrium Green's functions, changing sign at zero frequency as a result of thermal occupation, can be violated in driven-dissipative systems and it is in general not directly constrained by the structure of the stationary state density matrix. 

We have then applied our approach to the case of a Kerr nonlinear oscillator with incoherent driving and two-particle losses. Such a model turns out to be a perfect case study, since the properties of its stationary density matrix are well known, while its spectral features reveal a number of surprises. In particular the resonator density of state shows a strong dependence from the strength of the Kerr nonlinearity, a feature completely absent in the steady state populations only set by pump/loss ratio. Even more interestingly, in  the regime of large interaction and large non-equilibrium imbalance a NDoS emerges, an effect which would not be possible in thermal equilibrium.

We have summarized the behavior of the spectral function of this model in the phase diagram of Figure~(\ref{fig:mapZeroSpecFunc}) which shows that NDoS is not necessarily related to an inverted population in the steady state density matrix. In order to build physical intuition and to better understand the origin of this result we have developed a semi-analytical approach that starts from the spectral function of the isolated problem and adds a lifetime due to dissipation  in the spirit of a Fermi Golden Rule. This method, which turns out to be equivalent to a perturbation theory in the dissipation where only the eigenvalues of the Liouvillian are corrected, was able to partially capture the NDoS effect, at least for sufficiently large interaction and whenever the stationary density matrix shows population inversion. Finally we have shown that including the perturbative correction to the eigenstates of the Liouvillian results into a new mechanism for NDoS, due to the emergence of complex weights in the spectral function. This turns to be crucial to capture NDoS in the regime where the populations of the steady state are not yet inverted.

To conclude we mention that the  approach outlined here is rather general and can be used to shed light on the spectral properties of other small driven-dissipative quantum models. Interesting future directions include for example the study of resonance fluorescence lineshapes beyond the two-level system limit~\cite{BaurEtAlPRL09,delValleLaussyPRL10}, the spectral features of a coherently driven cavity across a zero-dimensional dissipative phase transition~\cite{CarmichaelPRX15,FinkEtAlPRX17,casteelsFazioCiutiPRA2017} or applications related to quantum synchronization~\cite{LorchEtAlPRL17,NiggPRA18}.

\emph{Acknowledgements.} This work was supported by the University of Chicago through a FACCTS grant (``France and Chicago Collaborating in The Sciences"), by the CNRS through the PICS-USA-147504 and by a grant ``Investissements d'Avenir" from LabEx PALM (ANR-10-LABX-0039-PALM).

%We have considered the Lehmann's rep of green's functions for open systems comparing it to the equilibrium one. We have studied the simple case of a van der Pol oscillator for which there's a clear distinction between stationary state and dynamical properties and we have computed the retarded Green's function using the Lehmann's representation. We have shown some results for which non-equilibrium conditions and interactions are essential and we have used perturbation theory to get more insights on the Green's functions in this specific case. 

%\bibliography{boseHubbard.bib}
%%\bibliography{books.bib}
%%\bibliography{DMFT.bib}
%%\bibliography{LehmannRepOfNeqGreenFunctions.bib}
\bibliography{greenFunctionsOpen}

\end{document}